\documentclass{aa}  

\usepackage{graphicx}
\usepackage{txfonts}
\usepackage{mathtools}
\usepackage{amsmath}
\usepackage{pdfpages}
\usepackage{natbib}
\usepackage{url}
\usepackage{placeins}
\usepackage{tikz}
\usepackage{wasysym}
\usepackage{gensymb}
\usepackage{ulem}
\usetikzlibrary{shapes,arrows}

\newcommand{\eq}{eq. }

\newcommand{\sect}{Section }
\newcommand{\fig}{Figure }
\newcommand{\Fig}{Figure }
\newcommand{\tabl}{Table }

\newcommand{\Bv}{\textbf{B}}

\newcommand{\Vm}[1] {\langle{#1}\rangle}

\begin{document}

   \title{Numerical dependencies of the galactic dynamo in isolated galaxies with SPH}
   \author{Robert Wissing and Sijing Shen
          }
   \institute{
   Institute of Theoretical Astrophysics, University of Oslo, Postboks 1029, 0315 Oslo, Norway \\ \email{robertwi@astro.uio.no; sijing.shen@astro.uio.no;}}
   \date{}
 
  \abstract
   { 
   Simulating and evolving magnetic fields within global galaxy simulations provides a large tangled web of numerical complexity due to the vast amount of physical processes involved. Understanding the numerical dependencies that act on the galactic dynamo is a crucial step in determining what resolution and what conditions are required to properly capture the magnetic fields observed in galaxies.
   Here, we present an extensive study on the numerical dependencies of the galactic dynamo in isolated spiral galaxies using smoothed particle magnetohydrodynamics (SPMHD). We performed 53 isolated spiral galaxy simulations with different initial setups, feedback, resolution, Jeans floor and dissipation parameters. The results show a strong mean-field dynamo occurring in the spiral-arm region of the disk, likely produced by the classical alpha-omega dynamo or the recently described gravitational instability dynamo. The inclusion of feedback is seen to work in both a destructive and positive fashion for the amplification process. Destructive interference for the amplification occurs due to break down of filament structure in the disk, increase of turbulent diffusion and the ejection of magnetic flux from the central plane to the circumgalactic medium. The positive effect of feedback is the increase in vertical motions and the turbulent fountain flows that develop, showing a high dependence on the small-scale vertical structure and the numerical dissipation within the galaxy. Galaxies with an effective dynamo saturate their magnetic energy density at levels between 10-30\% of the thermal energy density. The density averaged numerical Prandtl number is found to be below unity throughout the galaxy for all our simulations, with an increasing value with radius. Assuming a turbulent injection length of 1 kpc, the numerical magnetic Reynolds number are within the range of $Re_{mag}=10-400$, indicating that some regions are below the levels required for the small-scale dynamo ($Re_{mag,crit}=30-2700$) to be active.
  }

   \maketitle
%

\section{Introduction}
Observations in the last few decades have revealed that many galaxies exhibit strong magnetic fields, with strengths from around several $\mu G$ for the Milky way and nearby galaxies \citep{2009Natur.462.1036O,2010ASPC..438..197F,2013Sci...341..147B,2015A&ARv..24....4B} up to several $mG$ in starburst galaxies \citep{2004A&A...417..541C,2011A&A...535A..79H,2013A&A...555A..23A,2008ApJ...680..981R}. The magnetic energy in these galaxies are found to be close to equipartition with the thermal and turbulent energies  \citep{1990ApJ...365..544B,1996ARA&A..34..155B,2009ApJ...702.1230T}, meaning that they are strong enough to dynamically affect the galaxy. Furthermore, it has been observed that the morphology of the magnetic field within disk galaxies exhibits a large-scale spiral structure \citep{2013pss5.book..641B}. In disk galaxies with a strong density wave structure, the magnetic field tightly coincides with the optical spiral arms, as in M51 and M83 with a strength of around $20-30\mu G$ \citep{2011MNRAS.412.2396F,2016A&A...585A..21F}. For galaxies with weaker density structure, the magnetic field can instead form large-scale magnetic arms not coinciding with the optical spiral arms, like in NGC6946 \citep{2007A&A...470..539B}. 
\\ \\
The strong magnetic fields observed can contribute a significant non-thermal pressure component to the galaxy, which can suppress star-formation rates and heavily affect the structure of the interstellar medium \citep{2013MNRAS.432..176P,2015MNRAS.447.3678B}. The correlation between star formation rate density and the magnetic field strength have been measured from observation to be between $B\propto SFR^{0.18}$ \citep{2007A&A...462..933C} and $B\propto SFR^{0.3}$ \citep{2014AJ....147..103H}. Within the ISM the magnetic field also plays an important role in the dynamics of molecular clouds, where strong fields can lead to more massive but fewer cloud cores \citep{2005ApJ...618..344V,2008MNRAS.385.1820P}. Another interesting aspect of magnetic fields within galaxies, is that they can suppresses the development of fluid instabilities \citep{1995ApJ...453..332J,2015MNRAS.449....2M}. This can allow for cold gas to survive longer within the predominately hot galactic outflows. This could provide a possible explanation to the observational significant component of cold molecular gas seen in galactic outflows \citep{2010AJ....140..445C,2014A&A...562A..21C,2015ApJ...814...83L,2018ApJ...856...61M}. The strength and structure of magnetic fields in galaxies also determine the transport of cosmic rays (CRs), which together with magnetic fields can efficiently drive galactic outflows \citep{2012MNRAS.423.2374U,2013ApJ...777L..16B,2016ApJ...824L..30P,2018ApJ...868..108B}. 
\\ \\
The magnetic fields in galaxies are thought to originate from weak initial seed fields that are in time rapidly amplified by the galactic dynamo. There are several possible origins for the initial seed field, such as through the Biermann battery \citep{2005ApJ...633..941H}, from shock/ionization fronts \citep{1994MNRAS.271L..15S}, from plasma instabilities \citep{1973SvA....17..137B,2005LNP...664....1R,2009ApJ...693.1133L,2012PhRvL.109z1101S,2013ApJ...778...39S}, from a primordial origin \citep{2013A&ARv..21...62D}. Additional seed fields can also be injected to the ISM through stellar winds, supernova and AGN feedback. These initial seed fields can potentially be very weak, for example the Biermann battery process is estimated to generate fields of the order of $10^{-20}G$. This would require the galactic dynamo to amplify the seed field by more than 14 orders of magnitude to replicate the current observed magnetic field strength of nearby galaxies. In addition, observations indicate that strong fields were already in place at high redshift \citep{2002RvMP...74..775W,2008Natur.454..302B}. The turbulent medium of galaxies give rise two distinct groups of dynamo processes that can achieve these sort of growth rates. The first is the small-scale/fluctuating dynamo, which can occur in any turbulent system as it is driven by the random stretching, twisting and folding of the field lines. This produces randomly orientated magnetic fields at scales smaller than the driving scale of turbulence. As the most rapid stretching, twisting and folding happens on small-scales, the e-folding time will be set by the turnover time of the viscous-scale eddies, giving analytical e-folding times predicted to be less than $10-100$ Myr for galaxies \citep{2004ApJ...612..276S}. The small-scale dynamo will eventually saturate when the field becomes sufficiently strong to back-react on the flow and hinder the twisting of the field. The other group of dynamo processes are the mean-field dynamos, which generates magnetic fields at higher scales than the driving scale of turbulence. This requires that the underlying turbulence interact with larger scale inhomogeneities in the density or flow structure, for example with shearing flows and density stratification. This causes larger scale polarities to appear in the magnetic field.
\\ \\
The turbulence in the ISM is continuously regenerated by various stirring mechanisms, which include supernova explosions and stellar winds \citep{1989ApJ...345..782M,2004ApJ...617..339B,2005A&A...436..585D,2006ApJ...653..361K,2009ApJ...694L..26G,2009SSRv..143..263B,2011ApJ...729...72P,2012ApJ...752..146L}, gravitational collapse and accretion \citep{1953ApJ...118..513H,2010A&A...520A..17K,2010ApJ...712..294E,2010ApJ...715.1302V,2011ApJ...731...62F,2012ApJ...750L..31R}, AGN feedback \citep{2016MNRAS.461..967M}, spiral-arm compression \citep{2008MNRAS.385.1893D,2008MNRAS.391..844D}, cloud–cloud collisions \citep{2009ApJ...700..358T,2013ApJ...776...23B}, the magneto-rotational instability  \citep{2007ApJ...663..183P,2009AJ....137.4424T} and the galactic shearing flows. This makes the small-scale dynamo likely be active within the ISM. However, the efficiency and saturation of the small-scale dynamo will depend heavily on the fluid parameters (Reynolds number, Magnetic Reynolds number, Prandtl number and Mach number) and the mixtures of solenoidal-to-compressible modes within the turbulent flow \citep{2014ApJ...797L..19F}.
\\ \\
Turbulence can be decomposed into two modes, compressive (potential) modes ($\nabla \times v=0$) and solenoidal (rotational) modes ($\nabla\cdot v=0$). Different turbulence driving mechanisms can excite more or less of either mode. Feedback processes such as supernova explosions and gravitational collapse are compressive drivers and mainly inject compressive modes within the fluid, while the magneto-rotational instability and shearing flows are solenoidal drivers that mainly inject solenoidal modes. While these drivers initially inject a certain mode, each mode can feed on the other as they interact with their environment \citep{1973PASJ...25....1S}. This is important as turbulence containing only compressive modes cannot directly excite the small-scale dynamo as it does not impart any vorticity to the fluid (thereby no twisting and folding of the field lines) \citep{2006MNRAS.370..415M}. However, indirectly through significant transfer between compressional modes to solenoidal modes the dynamo can be excited, for example through non-linear interactions of colliding shocks \citep{1994ApJ...428..186V,2003JFM...478..237S,2007ApJ...665..416K,2010A&A...512A..81F}, rotation and shear forces \citep{2011A&A...528A.145D}, baroclinicity \citep{2016ApJ...822...11P} and through viscous forces \citep{2006MNRAS.370..415M,2011PhRvL.107k4504F}. The developed/saturated ratio of solenoidal to compressional modes is thus a complicated matter which will depend heavily on the fluid conditions and environment.
\\ \\
Take for example the turbulence in the ISM, which is highly supersonic at Mach numbers of around 10–100 \citep{2004RvMP...76..125M}. This environment generates significant energy transfer between compressional and solenoidal modes, and can significantly affect the dynamo process. This works both ways, with coherent vortex structures being destroyed by the formation of shocks \citep{2004PhRvE..70a6308H}, and vorticity being generated in the interaction of oblique colliding shocks \citep{2003JFM...478..237S,2007ApJ...665..416K}. As shown by \cite{2010A&A...512A..81F}, the balance between these two processes is highly dependent on the mach number, where at higher mach numbers vorticity generation was shown to become the dominant factor.
\\ \\
In the case of numerical simulations we can potentially fail to resolve the energy transfer between modes due to resolution constraints. An interesting example illustrating this is the work by \cite{2011ApJ...731...62F} and \cite{2012ApJ...745..154T}, which showed that to properly resolve the ratio of solenoidal to compressional turbulence generated in gravitional collapse (roughly $E_{sol}/E_{tot}=2/3$), the local Jeans length were required to be resolved by a high enough number of resolution elements (30 in \cite{2011ApJ...731...62F} and 64 in \cite{2012ApJ...745..154T}). This is due to the compressional turbulence being injected at the Jeans length is converted into solenoidal motions at smaller scales which needs to be resolved. For the small-scale dynamo this is significant, as the magnetic field amplification below this resolution criteria showed no growth or a reduction in the field strength (relative to the spherical adiabatic compression of the field lines $B\propto\rho^{2/3}$). It is still somewhat uncertain how resolution and different numerical schemes affect the transfer mechanisms mentioned above.
\\ \\
Another big factor for the dynamo relates to the injection length of these turbulent drivers, as this is the scale that, together with the velocity and dissipation parameters, determine what the efficient Reynolds numbers and magnetic Reynolds number are in the simulation
\begin{equation}
Re = \frac{L_{inj}\sigma_v}{\nu}, 
\label{eq:re}
\end{equation}
\begin{equation}
Re_{mag} = \frac{L_{inj}\sigma_v}{\eta},
\label{eq:rem}
\end{equation}
here $L_{inj}$ is the driving/injection scale, $\sigma_v$ is the turbulent velocity dispersion on that length-scale, $\nu$ is the viscosity coefficient and $\eta$ the resistivity coefficient. The growth of the turbulent dynamo depend strongly on these two parameters, where higher numbers generally lead to faster growth of the magnetic field. The so-called critical magnetic Reynolds number can be seen to represent the minimum separation required between the driving scale and the dissipative scales to drive the small-scale dynamo. The critical magnetic Reynolds number depends on both the fluid environment (shear, rotation, compressibility) and the Reynolds number, and remains fairly uncertain. Values of around $Re_{mag} = 30-2700$ is given from different model analysis and numerical simulations of turbulent systems \citep{2004MNRAS.353..947H,2004MNRAS.353..947H,2005PhR...417....1B,2012ApJ...754...99S,2014ApJ...797L..19F}. While the range of values are fairly large, there is a tendency for higher critical magnetic Reynolds number for higher compressibility. The different turbulence drivers that we discussed earlier will all have different injection lengths, some more global (e.g., spiral-arm compression and shear) and others more local (e.g., supernova and AGN feedback, collapse). The turbulence injection length of supernova feedback in simulation will heavily depend on the subgrid model and the environment in the ISM. This remains fairly unexplored for the SPH subgrid models, so it is hard to give a good estimate. High resolution ISM simulations with grid codes have shown that SNe have injection scales of roughly 60-200 pc \citep{2006ApJ...653.1266J,2007ApJ...665L..35D,2013MNRAS.432.1396G,2017ApJ...850....4H}, which include simulations with and without SN clustering. This is smaller than the average size of superbubbles from SN clustering, which lies between 0.5 to 1 kpc. The reason given for the small injection scale is that local breakup of bubbles occur earlier in the ISM due to the strong density and pressure gradients present. Still, this is using high resolution local simulations, and it is not necessary that this correlates to the same scale for our subgrid models of stellar feedback. Turbulent injection from spiral arm compression/break down will on the other hand occur on much larger scales (1-3 kpc). The small-scale dynamo of the ISM is predicted to saturate the magnetic field with energies at around $1-50$ percent of equipartition with the turbulent energy \citep{2005ApJ...625L.115S}.
\\ \\
While the small-scale dynamo gives a mechanism that can quickly amplify the field, it cannot produce the large-scale fields observed in galaxies. For this we need some sort of mean-field dynamo. The most well known mean-field dynamo is the classical $\alpha\Omega$ dynamo, which depends on the shear of the flow ($\Omega$-effect) and the small-scale velocity helicities in the flow ($\alpha$-effect). However, the $\alpha\Omega$ dynamo is plagued by the so called "catastrophic-quenching effect"\footnote{Occurs due to a build up of small-scale current helicities within the flow that can act to oppose the $\alpha$-effect.}, that effectively limits the saturation strength of the large-scale magnetic fields. The saturation can be shown to be proportional to $1/Re_{mag}$, which for the ISM $Re_{mag}\approx 10^{15}$ will result in very low saturation levels \citep{2002ApJ...579..359B,2005PhR...417....1B}. This is only true if one assumes that the helicity within active dynamo regions is conserved (closed boundary). However, in galaxies there are plenty of processes that can remove helicity from the disk and avoid quenching \citep{2004A&A...427...13B}. The growth rate of the $\alpha\Omega$-effect will strongly depend on both the global properties of the disk (scale height, shear parameter, etc.) and the small-scale properties (injection length, dissipation etc). Similarly, there is also the $\alpha^2$ dynamo, which solely relies on the small-scale velocity helicities in the flow to generate its mean-field and may be an important process in generating mean-fields in galaxies with more uniform rotation \citep{2005PhR...417....1B}.
\\ \\
Another type of mean-field dynamo that has recently emerged as a very interesting prospect for dynamo growth within astrophysical disks, is the gravitational-instability (GI) dynamo \citep{2019MNRAS.482.3989R}. This is a dynamo that is sustained by the gravito-turbulence injected during spiral arm compression, which generates vertical rotating flow rolls. During compression the toroidal field is pinched, lifted and folded by these flow rolls, generating new radial fields. These radial fields are then sheared by the differential rotation generating toroidal fields, closing the dynamo loop. This is similar to the $\alpha\Omega$ type dynamo, but slightly different as it is governed by larger scale motions than the turbulent helical motions. The growth rate of this dynamo depend strongly on the cooling rate and the effective Reynolds number. The critical Reynolds number has been shown to be around the order of unity ($Re_{mag,crit}=4$ for $\tau_c=20\Omega^{-1}$, where $\tau_c$ is the cooling time and $\Omega$ the rotation rate). Above this value the dynamo starts to saturate close to quasi-equipartition with the turbulent energy, but shows a decrease in saturation above $Re_{mag}\ge100$. In \cite{2019MNRAS.482.3989R}, high $Re_{mag}\ge100$ simulations showed increased small-scale structure in the large-scale magnetic ropes inside spiral waves. It was suggested that the small-scale fields was generated by the turbulent small-scale dynamo which can act to break down the large-scale field through parasitic (secondary) instabilities. This would be similar to the break down of channel modes for the MRI and reminiscent of the "catastrophic-quenching effect" for the $\alpha\Omega$-effect. There has however, as of yet not been an extensive study of higher $Re_{mag}$ simulations looking at this dynamo, making its behaviour in this regime difficult to predict.
\\ \\
Apart from the $\alpha\Omega$-effect and the GI-dynamo, there are several other proposed mechanisms that can develop coherent large-scale magnetic fields in galaxies, for example cosmic-ray driven dynamo  \citep{2003A&A...401..809L,2009ApJ...706L.155H}, shear-current effect \citep{2003PhRvE..68c6301R,2004PhRvE..70d6310R,2015PhRvL.114h5002S} and the stochastic alpha-effect \citep{1997ApJ...475..263V,2000A&A...364..339S,2011PhRvL.107y5004H}. However, it remains unexplored if any of these additional mechanisms can lead to a sustained dynamo in a more realistic environment with global and open boundary conditions.
\\ \\
A natural way to study both the small-scale and large-scale dynamo in galaxies is through numerical simulations. There has been great advances in the understanding and numerical modeling of galaxies through the inclusion of physical processes such as the gravitational interaction between dark matter, stars and gas \citep{1980ApJ...236...43A,2001PhDT........21S,2002JCoPh.179...27D}; the hydrodynamic modeling of gas \citep{2002A&A...385..337T,2004NewA....9..137W,2010MNRAS.401..791S}; the formation of stars \citep{1992ApJ...391..502K}; the feedback/output from supernova and stellar winds \citep{1992ApJ...391..502K,2003MNRAS.339..289S}; the feedback/output from black holes \citep{2005Natur.433..604D}; the radiative cooling of gas \citep{2003MNRAS.345..561M,2010MNRAS.407.1581S}. With these physics included, cosmological simulations are able to reproduce many observables in galaxies. However, magnetic fields still remain one of the most often neglected parts, mostly due to the complexity and technical difficulties associated with them. As we have seen, the magnetic field is closely interwoven with the dynamical state of the galaxy. The environment will determine the growth of the magnetic field, which in turn will react back on the dynamics. This means that there can be strong dependencies between the magnetic field and the other different physical processes (e.g., star formation, supernova explosion, radiation transport, active galactic nuclei). These strong dependencies become clear when we look at previous simulations of galaxies with magnetic fields, which have shown to produce a wide range of different magnetic field amplification and saturations. Some show magnetic fields growing up to levels near equipartition with the turbulent energy \citep{2009ApJ...696...96W,2013MNRAS.432..176P,2017ApJ...843..113B}, while others end up with a relatively weak saturated field \citep{2017MNRAS.471.2674R,2018MNRAS.473L.111S,2018MNRAS.479.3343M}
\\ \\
There are also numerical difficulties to consider when modeling the magnetic field within galaxies. First of all, in numerical simulations, the smallest spatial scale that we resolve is restricted by the number of resolution elements that we can afford to use in the computation. This is clearly relevant for magnetic fields, as the dynamo processes mentioned above are heavily dependent on the small-scale dissipation of the system. Apart from the independent dissipation of the magnetic and kinetic energies, amplification of the magnetic field have also been shown to be heavily dependent on the ratio between the magnetic and kinetic dissipation, otherwise known as the magnetic Prandtl number. This ratio is often overlooked in numerical simulations, but is crucial to understand the amplification and saturation of the magnetic field \citep{2004ApJ...612..276S,2005PhR...417....1B,2021arXiv210501091W}. In nature, galaxies are expected to have magnetic Prandtl numbers far greater than unity (in molecular clouds $P_m\approx10^{10}$ \citep{2016JPlPh..82f5301F}). In numerical simulations, the numerical Prandtl number is set by the numerical dissipation scheme used for the velocity and magnetic fields. For grid code the Prandtl number remains fairly constant at around $P_m=2$ \citep{2007A&A...476.1123F,2007MNRAS.381..319L,2009ApJ...690..974S,2011PhRvL.107k4504F}, though these estimate are taken for subsonic flow and might change for supersonic flows. For SPH this ratio is more resolution dependent as seen in \cite{2021arXiv210501091W} and \cite{2016MNRAS.461.1260T} and lies somewhere between  $P_m=1-2$ for subsonic flow\footnote{Given the resolution in those papers and the default values of the numerical dissipation coefficients used for those codes.}. 
\\ \\
Another technical consideration involves the generation of unphysical divergence errors (magnetic monopoles), due to truncation errors in the numerical discretization and integration of the MHD equations. Large divergence errors can lead to both force errors and amplification errors for the magnetic field. It is therefore crucial to try to keep the divergence error as close to zero as possible. Galaxy simulations prove to be one of the more difficult simulations in regards to withholding the divergence constraint, due to the supersonic environment, shear, open boundaries and the large amount of subgrid recipes. However, in the last few decades there have been tremendous improvements in reducing and handling these errors within numerical simulations. In Eulerian codes the divergence-free constraint can be enforced to machine precision with the constrained transport method \citep{1988ApJ...332..659E}. This is not easily applicable for Lagrangian codes\footnote{\cite{2016MNRAS.463..477M} have shown that an implementation of constrained transport scheme with moving meshes is possible, though being limited to global time-stepping.}. However, improved divergence cleaning methods have been developed that can significantly reduce the error for SPH \citep{2012JCoPh.231.7214T,2016JCoPh.322..326T}.
\\ \\
In this paper, we will study in detail how different numerical parameters such as supernova feedback, resolution, Jeans floor, diffusion parameters and initial conditions affects the growth and saturation of the magnetic field. 
\\ \\
This paper is organized as follows. In Section~\ref{sec:theory}, we go through the simulation setup and the post-process analysis. In Section~\ref{sec:results}, we present our result. In Section~\ref{sec:discussion}, we discuss our results and present some concluding remarks.
\section{Simulation description}
\label{sec:theory}
\subsection{Simulation setup}
\label{sec:simsetup}
For all our simulations we use the MHD version of {\sc Gasoline2} with the same default set of code parameters as in \cite{2020A&A...638A.140W}, except number of smoothing neighbours set to $N_{neigh}=64$. Gasoline2 applies different gradient operators compared to traditional SPH (TSPH)\footnote{By traditional SPH we mean the MHD equations that are derived directly from the Euler-Lagrange equations with the traditional SPH density estimate $\rho_a=\sum_b m_b W_{ab}$. See \cite{2012JCoPh.231..759P} for more information.} which has shown to improve solutions near density discontinuities \citep{2017MNRAS.471.2357W}. In particular, when applied to the magnetized cloud collapse, jet formation was captured at lower resolutions and for weaker magnetic fields compared to previous SPH+MHD schemes \citep{2020A&A...638A.140W}. Additionally, GDSPH was able to successfully capture the development of the magnetorotational instability in a stratified medium, whereas previous meshless methods either developed numerical instability or saw decay of the turbulence after a short period of time  \citep{2019ApJS..241...26D,2021arXiv210501091W}. The non-MHD version of Gasoline2 has moreover been widely used to study galaxy formation in large-scale cosmological boxes, cosmological zoom-in simulations and isolated galaxy simulations.
\\ \\
For our initial conditions (IC) we use the isolated disk galaxy from the AGORA comparison project \citep{2014ApJS..210...14K}, which was modeled to be similar to a Milky-Way type galaxy at $z=0$. The IC was generated by the {\sc Makedisk} code \citep{2005MNRAS.361..776S}, which distributes the particles provided an equilibrium solution to the Jeans equation for a multi-component system including the halo, disk and bulge. The initial gas metallicity is set to solar values. This IC is very useful as it has been readily used in the literature, which allows for more comparison to our simulations. The AGORA comparison project offers several resolutions of this disk galaxy, for our simulations we split all the particles of the lowest resolution AGORA IC by 8 times (low resolution), 64 times (medium resolution) and 512 (high resolution). These are all relaxed with MHD, feedback and star formation all turned off but with cooling turned on together with a very high Jeans floor, to generate a similar smooth IC in each case. The number of particles together with the mass and length resolutions for the three ICs are given in \tabl \ref{table:1}.
\begin{table*}[!h]
\centering
\begin{tabular}{lrrrrrrrrrrc}
\hline\hline  
\smallskip
$ $&$N_{tot}$&$m_{gas}[M_\odot]$&$m_{star}[M_\odot]$&$m_{dark}[M_\odot]$&$\epsilon_{gas}[kpc]$\\
\hline
Low&$2.5\cdot10^6$&$1.0\cdot10^4$&$1.0\cdot10^4$&$1.6\cdot10^6$&0.02 \\
Medium&$2.0\cdot10^7$&$1.3\cdot10^3$&$1.3\cdot10^3$&$2.0\cdot10^5$&0.01 \\
High&$1.6\cdot10^8$&$1.7\cdot10^2$&$1.7\cdot10^2$&$2.5\cdot10^4$&0.005 \\
\hline
\hline
\end{tabular}
\caption{The three initial conditions used in our simulations. Stating the number of particles ($N_{tot}$), initial particle masses ($m_{gas},m_{star},m_{dark}$) and the softening length of the gas ($\epsilon_{gas}$). $m_{star}$ refers to the mass of star particles that form during the simulation. In addition, there is an old stellar disk component in the IC, which consists of star particles that have masses of around $m_{star,old}\approx 4.3 m_{star}$ }
\label{table:1}
\end{table*}

\subsection{Star formation and particle splitting}
The simulations include a uniform UV background radiation ($z=0$) \citep{1996ApJ...461...20H}, radiative heating and cooling due to hydrogen, helium and metals \citep{2010MNRAS.407.1581S}, and photoelectric heating. The diffusion of metals and thermal energy are modeled using a subgrid turbulent mixing model \citep{2008MNRAS.387..427W,2010MNRAS.407.1581S}. We use a stochastic star formation recipe for our simulations, which is based on the models developed by \cite{1992ApJ...391..502K} and \cite{2006MNRAS.373.1074S}. Gas particles are eligible to become stars if they are in a converging flow with a density that is above the density threshold of $\rho_{SF}>100$ and with a temperature that is below $T<10^4K$. Each timestep there is a probability that a star particle will form. This probability is based on the theoretical star formation rate, which we can get from the Schmidt law:
\begin{equation}
    \dot{\rho_*}=\epsilon_{SF}\frac{\rho_{gas}}{t_{ff}}
\end{equation}
Here $\epsilon_{SF}=0.05$ is the star formation efficiency and $t_{ff}=\frac{1}{\sqrt{4 \pi G \rho_{SF}}}$ is the free fall time. The probability of forming a star particle each time step is then given by:
\begin{equation}
    P_{SF}=1-\exp{\left(-\frac{\epsilon_{SF} \Delta t }{t_{ff}}\right)}
\end{equation}
Here $\Delta t$ is the length of the current timestep. When a star particle forms, it replaces the whole gas particle, this means that every star formation event will leave holes in the magnetic field. This can be seen as the magnetic flux getting trapped within the star particle. However, this does affect both the local magnetic energy and the local divergence error. Star particles that form within the simulation represent a stellar population that evolves according to stellar theory, with an initial mass function following \cite{2003PASP..115..763C}. Feedback from stellar winds, Type Ia and Type II SNe eject mass and metals into the ISM, which reduces the mass of the star particle and increases the mass of the surrounding gas particles. If left unchecked, this can lead to situations where gas particles have significantly different masses. This is undesirable as the accuracy of the SPH method can quickly be degraded if there is a significant difference in particle masses within the smoothing kernel. A simple way to avoid this is to split the gas particles if they exceed 2 times their initial mass into two equal mass particles with the same properties, which are placed randomly within the original particles smoothing radius. This is the default way to handle it in {\sc Gasoline2} but can lead to more grid noise and divergence errors for the magnetic field. To reduce this error, we instead present a new particle-splitting method.

The main issue with distributing the child particles within the smoothing kernel is that neighbors strongly feel the change of the particle split. In addition, because the distribution is random it can place the child particles close to other neighboring particles or across density gradients, leading to spurious fluctuations in the local field. These effects can be mitigated by instead splitting the particle within the interparticle distance ($h_{int}$) instead. This is similar to the methods presented in \cite{2006ApJS..163..122M} (particles placed on vertices of a cube with cube length $\frac{1}{2}h_{int}$) and \cite{2015MNRAS.451.3955C} (particles placed within local Voronoi cell). In our prescription, the two daughter particles are distributed $\frac{1}{3}h_{int}$ away from the parent particle on a line that is orthogonal to the closest neighbor (located $h_{int}$ away). Making the two new daughter particles to be an equal distance away from the closest neighbor minimizes the effect of the split. When a particle now splits, the neighbors do not directly see any significant change as the position of the child particles is very close to the position of the parent particle.
\\ \\
All our galaxy simulations are run until the magnetic field growth has stabilized, usually around $t_{end} = 1-3$ Gyr. The resolution, SN feedback parameters, Jeans floor and the magnetic field configuration are varied in different runs. Below we outline more details about the parameters altered. 
\subsection{Jeans floor}
\label{sec:jeansfloor}
To avoid numerical fragmentation it is important to ensure that gas does not collapse beyond the resolvable Jeans length \citep{1997ApJ...489L.179T,1997MNRAS.288.1060B}. This can happen in galaxy simulations when the gas becomes cold and dense enough, such that its Jeans length is smaller than the resolution length of the simulation. To ensure this we add a non-thermal pressure floor to our simulation, based on the method described in \cite{2008ApJ...680.1083R}.
\begin{equation}
P_{min}=\frac{N_J^2 h^2 G \rho^2}{\pi \gamma}
\end{equation}
where $h$ is the smoothing length, $\rho$ is the gas density, and $\gamma = 5/3$ is the adiabatic index. This ensures that the Jeans length is resolved by $N_J$ smoothing lengths. The classic condition of \cite{1997ApJ...489L.179T} states that the Jeans length should be resolved by atleast 4 resolution length. However, the value required depends on several factors, such as numerical method, resolution, initial conditions and included physics, as there can be many factors that prevent the gas from entering a phase where its vulnerable to artificial fragmentation. The Jeans floor effectively sets the smallest collapsed length-scale in the simulation. The smallest collapsed length scale will thus be roughly $L_{coll}=\epsilon N_J$. When we go to higher resolution we decrease this length-scale to keep the same 4 resolution length condition as before. This of-course is generally a desirable trait as we go closer to resolving the real deal, however for comparison sake this can muddy the result. Keeping the smallest collapsed length-scale the same as we increase the resolution ensures that the turbulent driving scales remain similar. In turn, this increases the number of resolutions elements that resolves the minimum Jeans length. This can be done by introducing a scaling law \citep{2018MNRAS.478..302S},
\begin{equation}
   N_J=N_{J,low}(m_{gas}/m_{gas,low})^{-1/3}. 
\end{equation}
Here, the subscript $low$ represent the values given for the 'Low' resolution simulation (see \tabl\ref{table:1}). As we increase the resolution by splitting all the particles by $8$, the effective $N_J$ for each higher resolution simulation becomes $N_{J,medium}=2N_{J,low}$ and $N_{J,high}=4N_{J,low}$. While $N_J=4$ is the classical condition for avoiding artificial fragmentation, the magnetic dynamo from gravitational collapse give more stringent constraints. As mentioned briefly in the introduction, \cite{2011ApJ...731...62F} found that resolving the Jeans length with at least $30$ resolution elements is required to properly capture the solenoidal-compressible ratio during gravitational collapse. However, the Jeans floor simply sets the minimum local Jeans length to be resolved by $N_J$, the majority of scales within the simulation can fulfill the magnetic dynamo condition even if the smallest scale does not ($N_J<30$). Nevertheless, the compressible modes generated at the small scales can potentially act destructively on the magnetic field growth at those scales. Therefore, it is instructive to investigate a wide range of $N_J$ values for galaxy simulations to see its effect.
\subsection{Feedback models}
\label{sec:fbmodels}
To investigate the effect of SN feedback on the amplification of the magnetic field in galaxies, we run several simulations with varying feedback parameters. The first parameter that we vary in our simulation is the energy injected to the surrounding ISM $E_{SN}$ in SN events. The three strengths that we use in our simulations are $\epsilon_{SN}=10^{51}$ ergs, $E_{SN,low}=0.5  \epsilon_{SN}$, $E_{SN,high}=2.0 \epsilon_{SN}$. We also employ two different feedback models in our simulations, the blastwave model \citep{2006MNRAS.373.1074S} and the superbubble model \citep{2014MNRAS.442.3013K}. For the superbubble model we also vary the number of surrounding particles injected with energy during a feedback event ($N_{FB}=1,64,200$).
\\ \\
A wide range of different feedback models have been developed to tackle the lack of resolution in galaxy simulations to resolve the Sedov-Taylor phase of a SN explosion. Early SN feedback models simply relied on a direct thermal injection to the surrounding medium. The issue with this is that the thermal energy is quickly radiated away before it can do any work on the surrounding medium as should be the case for the resolved Sedov-Taylor phase \citep{1992ApJ...391..502K}. A way to combat this involves switching off the radiative cooling of gas that has received feedback energy, enforcing an adiabatic phase, for some length of time. This is the philosophy of the first feedback model we employ, the blastwave model/delayed cooling model \citep{2006MNRAS.373.1074S}.
\\ \\
However, the blastwave model does have some significant downfalls. First, star formation is clustered; new stars are spatially and temporally correlated, and feedback from their individual winds and SNe merge, thermalize and grow as a superbubble rather than a series of isolated SNe. Second, because superbubbles have both hot gas $>10^6$ K and sharp temperature gradients, thermal conduction is significant \citep{1977ApJ...218..377W}. The superbubble feedback model from \cite{2014MNRAS.442.3013K} represents a more realistic model when simulating supernova feedback from cluster of stars (which each star particle represent). This is done by introducing a separate cold and hot phase for each particle. The evolution of the superbubble is accurately captured with the help of thermal conduction and subgrid evaporation, which regulates the hot and cold phases without the need of a free parameter. This makes the model more insensitive to numerical resolution compared to the blastwave model described before.
\\ \\
In this paper we do not employ any magnetic field injection during feedback events. This is because it is highly non-trivial on how to properly inject magnetic field to the surrounding particle distribution. We leave this to be the topic of future work.
\subsection{Numerical diffusion}
\label{sec:numdiff}
To get the Reynolds number and magnetic Reynolds number we estimate the numerical dissipation from the equivalent physical dissipation equations. This is done by recording the energy lost due to the artificial dissipation terms. From the Navier-Stokes equation we can estimate the shear viscosity with:
\begin{equation}
\label{eq:shearvisc}
\nu_{AD}=\frac{\left(\frac{du}{dt}\right)_{AV}}{\frac{1}{2}\left(\frac{\partial v^i}{\partial x^j}+\frac{\partial v^j}{\partial x^i}\right)^2+(\nabla\cdot\mathbf{v})^2} \ .
\end{equation}{}
Here, we have assumed the fixed ratio between the bulk viscosity and the shear viscosity, which follows from the continuum limit derivation ($\zeta_{AV}=\frac{5}{3}\nu_{AV}$) \citep{2010MNRAS.405.1212L}. We estimate the physical resistivity from the Ohmic dissipation law:
\begin{equation}
\eta_{AD}=\frac{\rho}{J^2}\left(\frac{du}{dt}\right)_{AR} \ .
\end{equation}{}
Taking the ratio of the two equations then gives us the numerical Prandtl number:
\begin{equation}
P_{m,AD}=\frac{\nu_{AD}}{\eta_{AD}} \ .
\end{equation}{}
After estimating the local velocity dispersion and the injection length, the Reynolds number and magnetic Reynolds number can be estimated using \eq \ref{eq:re} and \eq \ref{eq:rem}.
\subsection{Magnetic field configuration}
\label{sec:magconf}
The strength and configuration of the initial magnetic field can play an important role in its subsequent development. The simple choice is to just apply a constant field parallel to one direction, for a galactic disk initiating it in the parallel direction of the angular momentum vector ($\hat{\textbf{z}}$) or orthogonal direction can be appropriate choices.
$$B_{init}=B_0 \hat{\textbf{z}}$$
$$B_{init}=B_0 \hat{\theta}$$
While not being a very realistic magnetic configuration for an evolved galaxy, it does present the system with a straight forward initial polarity which can subsequently effect the underlying field growth. The main issue with a constant field is that low density region can become very magnetically dominated to begin with. 
\\ \\
A more realistic magnetic configuration can be achieved by taking the flux freezing consideration into account, which would mean that the strength of the magnetic field would more closely follow the initial density distribution ($B \propto \rho^{2/3}$ for spherical collapse). To keep the initial magnetic field divergence-less while scaling with the density requires a more complex field configuration. The easiest way to construct such a field is with the use of a vector potential. However, the initial field will in this case be dependent on resolution and the accuracy of the gradient estimate. For the low resolution this can generate quite a noisy initial field at the free surfaces (due to noisy gradient estimates). Instead we construct a vertical and toroidal  density based field simply by:
\begin{equation}
\label{eq:vertfield}
B_{0,z} = B_0 \left(\frac{\rho}{\rho_0}\right)^{2/3} \hat{\textbf{z}}
\end{equation}
\begin{equation}
\label{eq:torfield}
B_{0,\theta} = B_0 \left(\frac{\rho}{\rho_0}\right)^{2/3} \hat{\theta}
\end{equation}
This magnetic field will not be divergence-less to begin with, but is statically cleaned using our divergence cleaning before the proper runs. To confirm the divergence-less constraint we track the normalized divergence error initially and during the simulation 
\begin{equation}
\label{eq:divBerr}
\epsilon_{divB}=\frac{h|\nabla\cdot \Bv|}{|B|} .
\end{equation}
During the simulation the mean of this quantity should preferably remain below $10^{-2}$ but higher values can still be acceptable, depending on the system. We also measure the normalized Maxwell stress:
\begin{equation}
    \alpha_{MW}=-2\frac{\Vm{B_RB_{\phi}}}{\Vm{B^2}}.
\end{equation}
\subsection{Analysis of turbulence}
\label{postanal}
There are several methods in which one can go about to define the turbulent velocity. In this paper we simply remove the mean rotation from the azimuthal component:
\begin{equation}
    v_{turb}^2=v_r^2+(v_{azi}-v_{rot})^2+v_z^2
\end{equation}
The rotation velocity can be estimated by calculating the $v_{circ}$ from the gravitational influence within the midplane or by remove the averaged cylindrical radial profile of $v_{azi}$ for the gas (taken over 0.15kpc radial bins). These give similar result and we use the latter method in this paper. Another popular way to estimate the turbulence is to calculate the velocity dispersion within the smoothing kernel. However, a negative of this method is that the length-scale at which the velocity dispersion is calculated will depend on the resolution and density. The effective turbulent kinetic pressure is given by:
\begin{equation}
    P_{vel}=\frac{\rho v_{turb}^2}{2}
\end{equation}
We also define the inverse of the thermal plasma beta ($\beta_{th}$) and the turbulent plasma beta ($\beta_{vel}$):
\begin{equation}
 \beta_{th}^{-1}= \frac{P_{mag}}{P_{th}}
\end{equation}
\begin{equation}
 \beta_{vel}^{-1}= \frac{P_{mag}}{P_{vel}}
\end{equation}
Together with the turbulent Mach number ($\mathit{M}$) and the shearing parameter ($q$).
\begin{equation}
    \mathit{M}=\frac{|v_{turb}|}{c_s}
\end{equation}
\begin{equation} q = -\frac{d\ln \Omega}{d\ln r} \ .
\end{equation}
Here $\Omega$ is the angular velocity and $c_s$ is the speed of sound.
After the simulation the particle data is interpolated to uniform grid data for post-analysis. To analyze the scale dependencies in the simulation we perform a spectral analysis of the velocity, magnetic and density fields using Fourier analysis. The resulting Fourier energy spectra is calculated using the spherical shell method from (e.g \cite{1995tlan.book.....F}):
\begin{equation}
\label{eq:power}
    E(k)dk=\int \widehat{A} \cdot \widehat{A}^* 2\pi k^2 dk
\end{equation}
Here $\widehat{A}$ and $\widehat{A}^*$ represents the Fourier transform and its conjugate of quantity $A$. The integration occurs over spherical shells in Fourier space with radius $k=\sqrt{k_x^2+k_y^2+k_z^2}$.
\\ \\
The compressive and solenoidal component of a given field ($A$) can be extracted using Helmholtz decomposition. Here, the Fourier transform is decomposed into an longitudunal and transverse component ($\widehat{A}=\widehat{A}_l+\widehat{A}_t$). The compressible part can be found by calculating: $\widehat{A}\cdot k = \widehat{A}_l\cdot k$ and then performing an inverse Fourier transform. This gives us $A_{comp}$ which can be removed from $A$ to give the estimated solenoidal component $A_{sol}$. This is then used to estimate the Fourier energy spectra for the solenoidal and compressive modes using \eq \ref{eq:power}. An interesting quantity to measure is the relative solenoidal to compresssive ratio at different scales.
\begin{equation}
\label{eq:solratio}
    E_{ratio}(k_{inj})=\frac{\int_{k_{inj}}^{\infty}E_{sol}(k)dk}{\int_{k_{inj}}^{\infty}E_{tot}(k)dk}
\end{equation}
In general, the pure velocity and magnetic energy scaling is investigated, where $A=v_{turb}$ and $A=v_{alf}=\sqrt{\frac{B^2}{\rho}}$. However, for the supersonic turbulence and the large range of scales that we cover in these simulations, it becomes more interesting to investigate the scaling dependencies of the velocity and magnetic densities, where $A=v_{turb}\sqrt{\rho}$ and $A=v_{alf}\sqrt{\rho}=B$.\footnote{Potentially one can also use the density scaling $\rho^{1/3}v_{turb}$ and $\rho^{1/3}v_{alf}$, which given that turbulence is saturated leads to the original \cite{1941DoSSR..30..301K} scaling for the kinetic turbulence. This is because within the inertial range, this density scaling ensures a constant energy flux \citep{2007ApJ...665..416K}.}
\section{Simulation results}
\label{sec:results}
The default initial magnetic field is set in to be in the vertical direction using \eq\ref{eq:vertfield} with $B_0=10^{-3}\mu G$ and $\rho_0=6.77331\cdot 10^{-23}$, this correlates to an initial central thermal plasma beta of $\beta_{0,centre}=10^7$. The artificial resistivity coefficient is set to $\alpha_B=0.5$ as the code default. For the simulations including feedback the default is the superbubble scheme with strength $1\epsilon_{SN}$ and number of injection particles set to $N_{FB}=64$. 
\subsection{Simulations with no feedback}
\label{sec:nofb}
\begin{figure*}[!ht]
    \centering
    \includegraphics[width=\hsize]{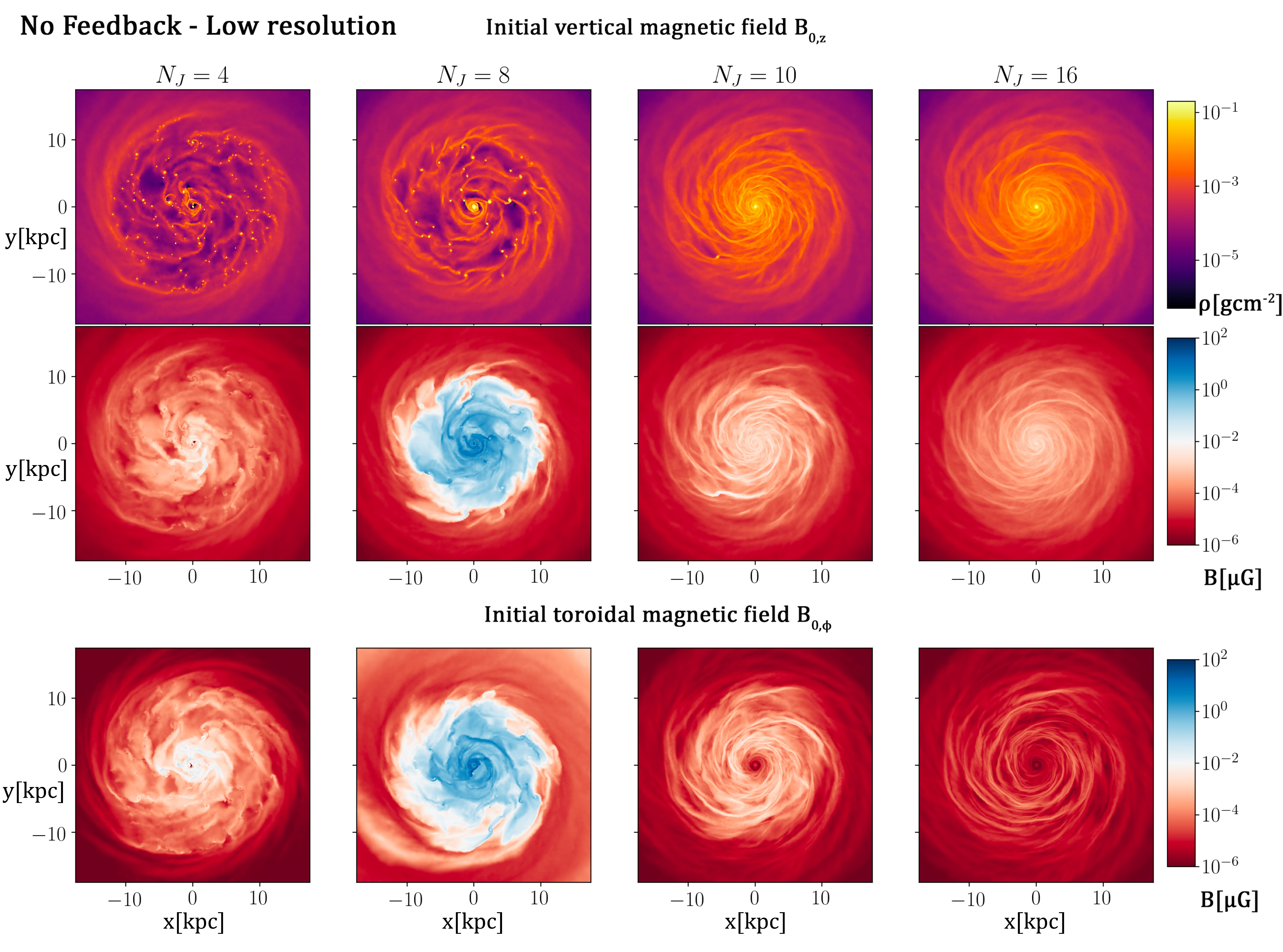}
    \caption{The face-on evolution of the no feedback simulations with varying Jeans floor and initial magnetic field geometry ($B_{0,z}$ two top panels, $B_{0,\phi}$ bottom panel). The time for each column/Jeans floor ($N_J=4,8,10,16$) is $t=(0.5,1,1,1)$ Gyr respectively. The $N_J=8$ case exhibit the strongest magnetic field growth at early times, likely through an active GI or $\alpha\Omega$ dynamo within the developed filamentary structure. Further reducing the Jeans floor leads to too much fragmentation of the disk, reducing the effect of the active dynamo. For higher Jeans floor 
    GI remain quite weak at this time, leading to either a weak amplification or decay of the magnetic field in the disk. However, in the case of the $N_J=10 $ and $B_{z,0}$ where a small fragment has started to form, the disk will become unstable at later time (see \fig \ref{fig:6}). 
    }
    \label{fig:1}
\end{figure*}
\begin{figure*}[!ht]
    \centering
    \includegraphics[width=\hsize]{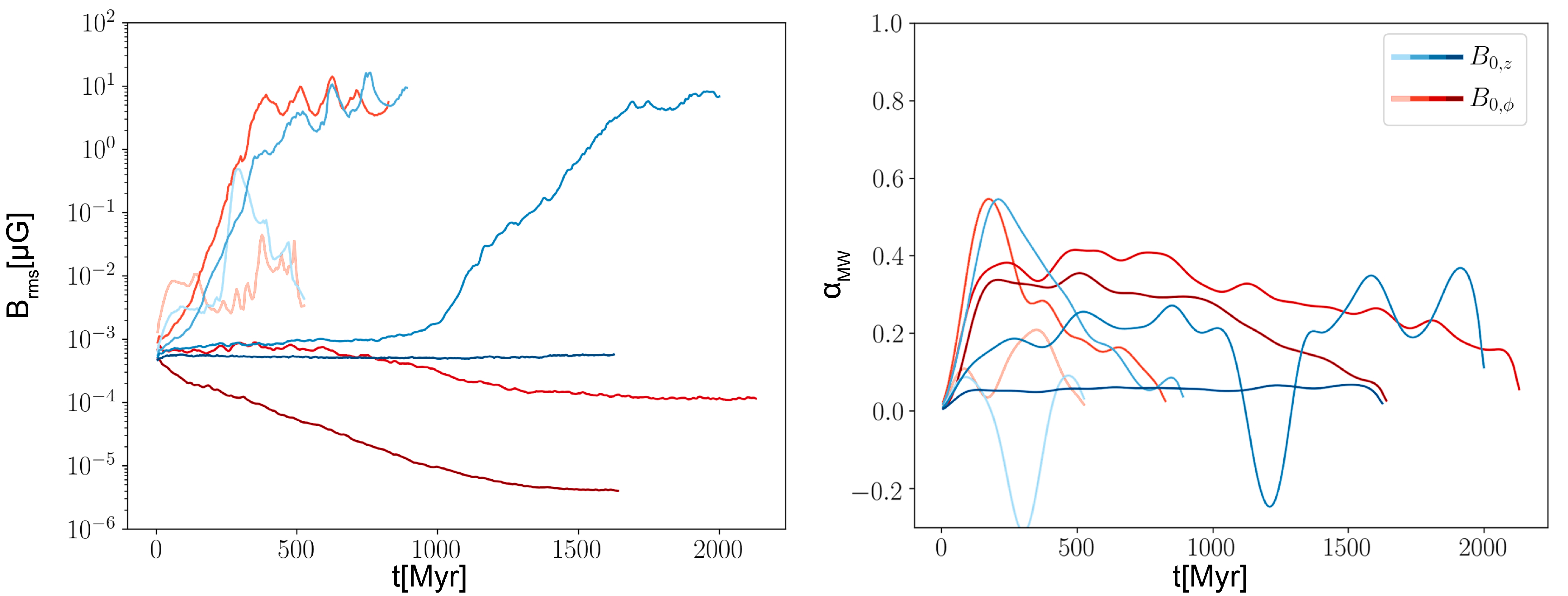}
    \caption{Time evolution of the average magnetic field strength (left panel) and the normalized Maxwell stress (right panel) for the no feedback runs. Blue lines represent simulations with an initial vertical field and the red lines represent simulations with an initial toroidal field. Darker colors represent higher Jeans floor ($N_J=(4,8,10,16)$). The $N_J=4$ and  $N_J=8$ simulations have a rapid amplification in the magnetic field early on due to spiral arm compression. We can see that we get the strongest $\alpha_{MW}$ exhibited during the growth phase of the $N_J=8$ cases, correlating to the generation of radial and toroidal fields during the dynamo process. The elevated level in $\alpha_{MW}$ for all the cases with an initial toroidal field is simply due to the initial field and normalization ($B_z\approx0$). }
    \label{fig:2}
\end{figure*}
\begin{figure}[!ht]
    \centering
    \includegraphics[width=\hsize]{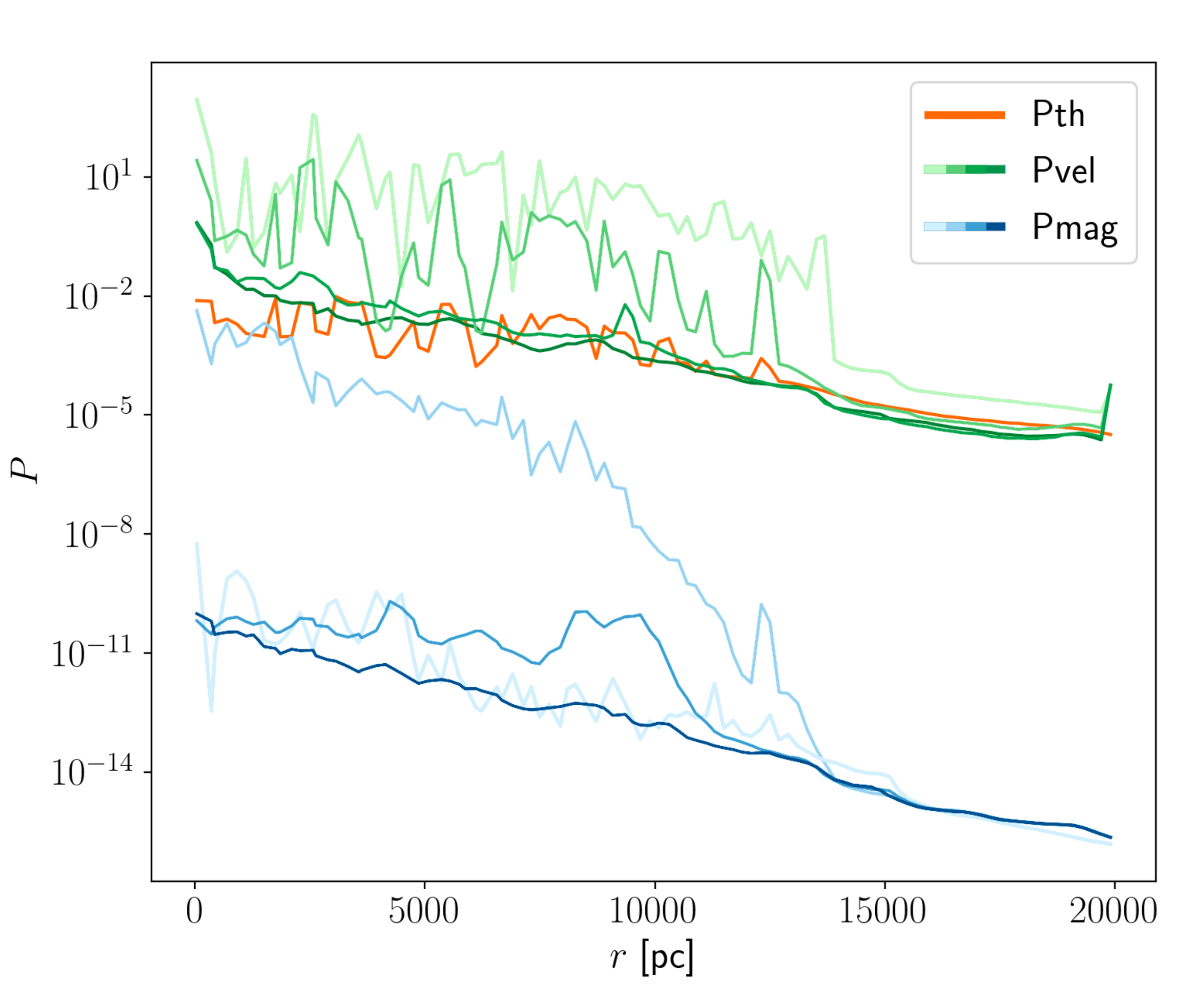}
    \caption{Radial profiles of the magnetic, turbulent and thermal pressures for the no feedback runs at around $t=1 \ \rm Gyr$. Darker colors represents higher Jeans floor ($N_J=(4,8,10,16)$). For the thermal pressure we only plot one line as it is similar across these simulations. For the $N_J=8$ case we can see that the magnetic pressure is of similar strength as the thermal pressure in the centre of disk. The magnetic pressure decreases with radius and returns to initial values beyond $14$ kpc. The turbulent pressure can be seen to significantly increases in both $N_J=4$ and $8$ as the disk is highly gravitationally unstable.}
    \label{fig:3}
\end{figure}
\begin{figure}
    \centering
    \includegraphics[width=\hsize]{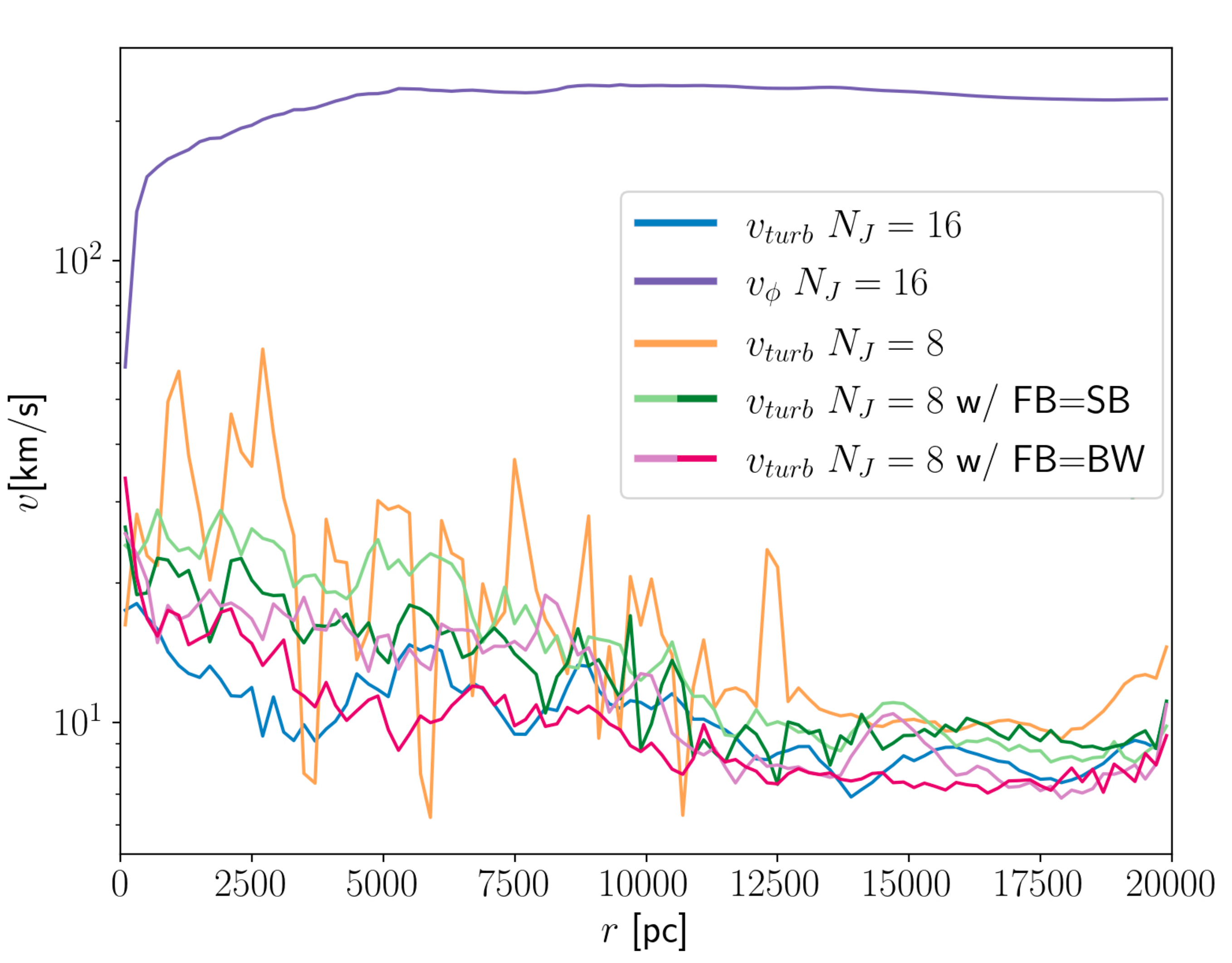}
    \includegraphics[width=\hsize]{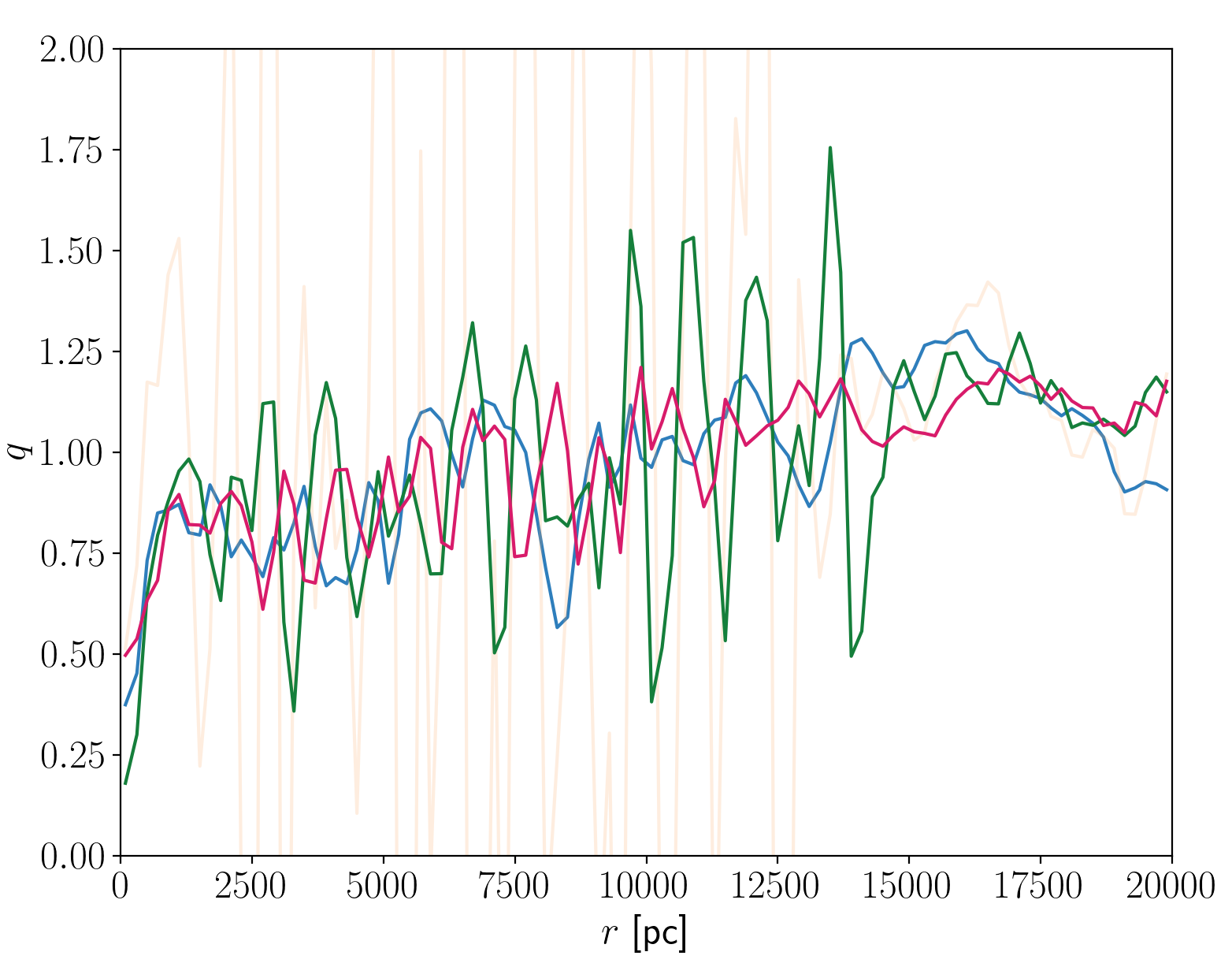}
    \includegraphics[width=\hsize]{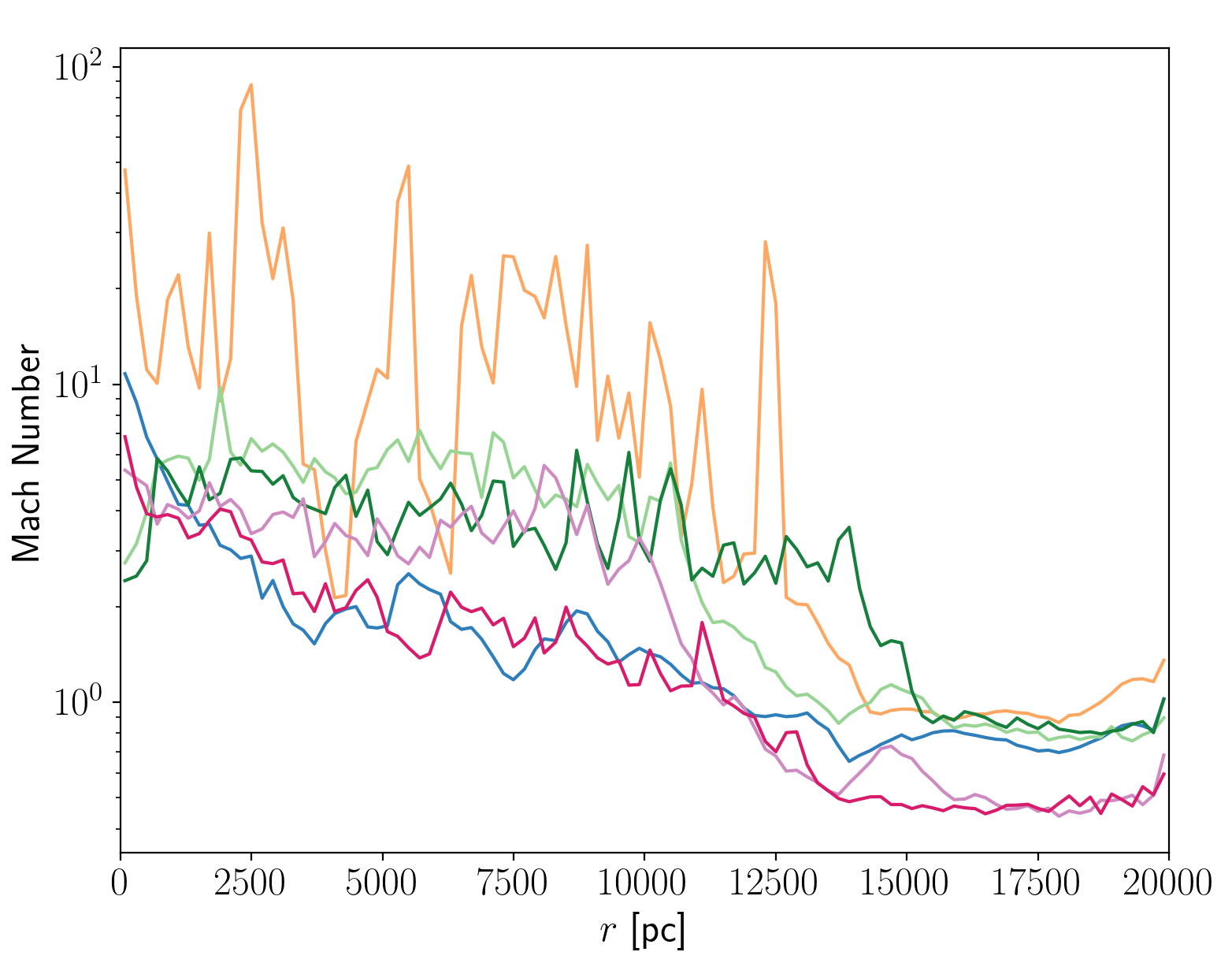}
    \caption{The radial profile of the velocity (top), the shear parameter (middle) and the Mach number (bottom) for the $N_J=8$ and $N_J=16$ no-feedback runs and for the SB and BW feedback runs with $N_J=8$ (light color $t=250$ Myr, dark color 2 Gyr). The turbulent velocities are enhanced by the gravitational instability of the disk and the inclusion of feedback. However, BW can be seen to produce a lower turbulent response compared to SB, especially at later times (within 10 kpc). In all runs we can see that there is a reduction of the shearing parameter towards the central region, reducing the effectiveness of an $\alpha\Omega$ type dynamo. Mach numbers are lower in the BW model due to the heating of the disk and the relatively low turbulent velocities.}
    \label{fig:4}
\end{figure}
\begin{figure}[!ht]
    \centering
    \includegraphics[width=\hsize]{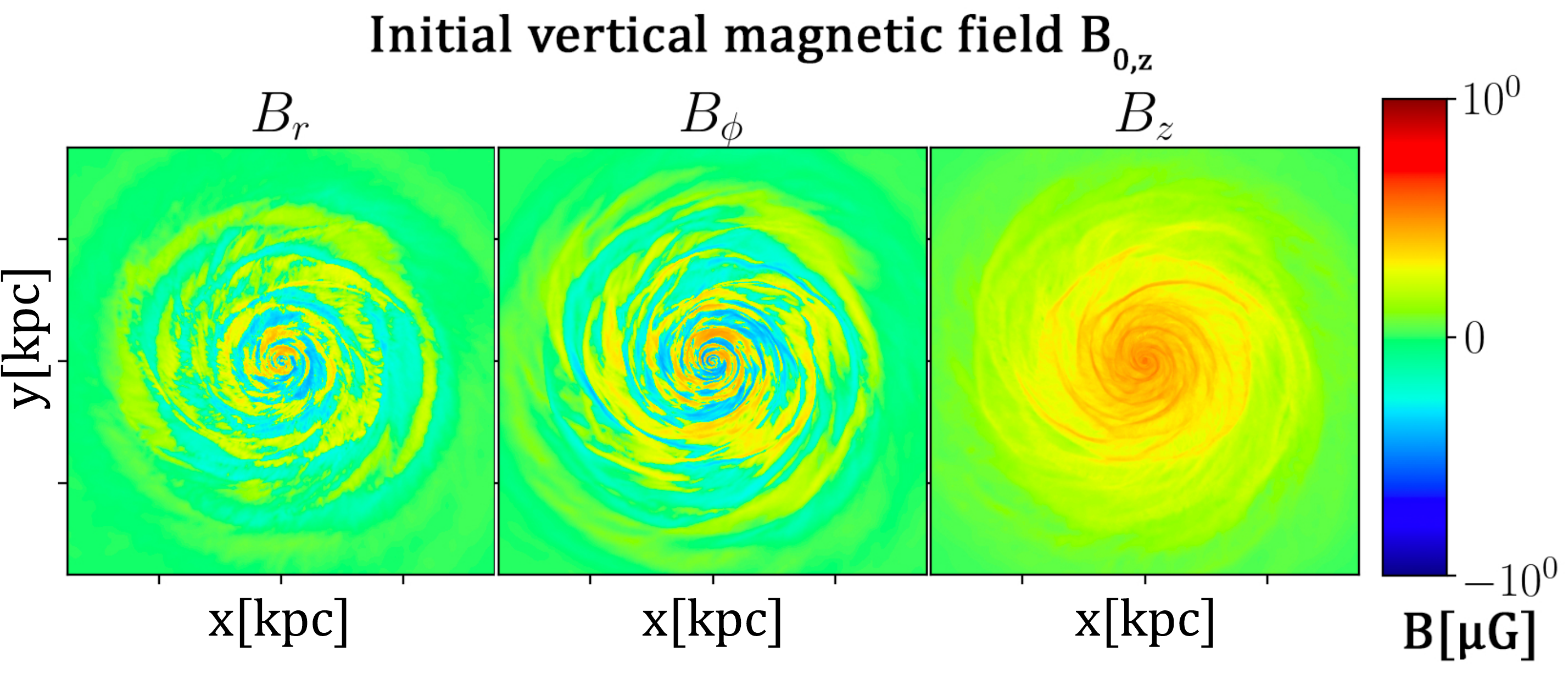}
    \caption{The magnetic field polarity of the $N_J=16$ no feedback simulation at $t=1$ Gyr with an initial vertical field. We can see field reversals throughout the disk in both the radial and toroidal directions. These are highly correlated to the spiral arm structure, as the reversals form together with the velocity perturbation induced by spiral shocks \citep{2016MNRAS.461.4482D}. The vertical field remains fairly correlated to the density and is similar in strength to the initial setup.}
    \label{fig:5}
\end{figure}
\begin{figure*}[!ht]
    \centering
    \includegraphics[width=\hsize]{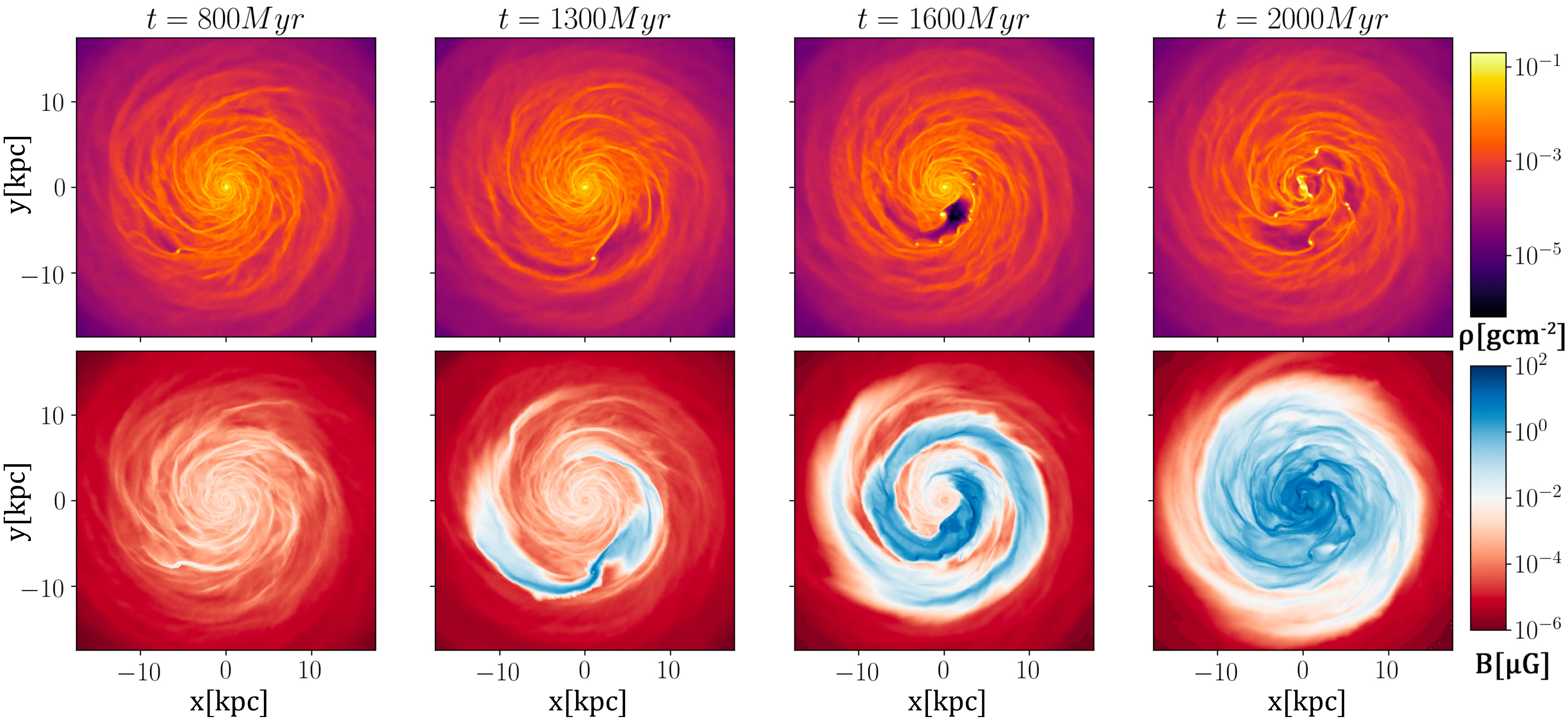}
    \caption{The development of a fragment within the disk for the $N_J=10$ $B_{0,z}$ case at later times ($t=800-2000 \ \rm Myr$), which subsequently destabilizes the disk and generates a strong magnetic field in the process. Top panel shows the density rendering and bottom panel the magnetic field strength.}
    \label{fig:6}
\end{figure*}
\begin{figure}[!ht]
    \centering
    \includegraphics[width=\hsize]{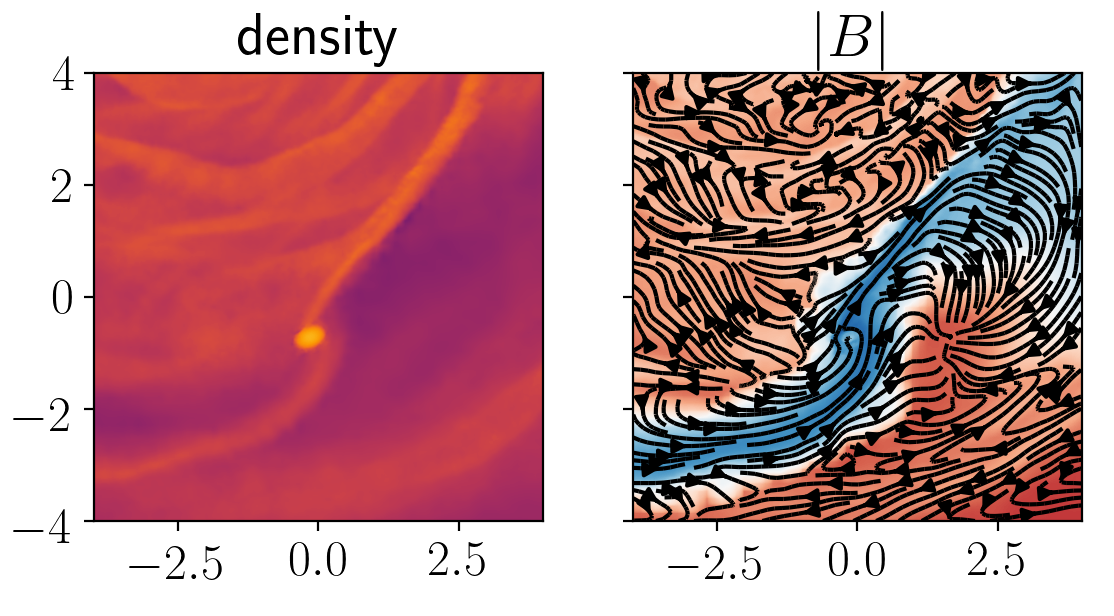}
    \caption{A closer look at the magnetic field structure around the fragment at $t=1300$ Myr. It is clear that the magnetic field traces the filamentary structure, with opposite/unstructured magnetic fields occurring in the low density region around it. This is similar to the magnetic field structure formed from the GI-dynamo simulations of \cite{2019MNRAS.482.3989R}, which due to magnetic flux redistribution generates opposite radial mean-fields within and outside the spiral arm. In addition, we can see that there is a strong field reversal in the gap in front of the fragment.}
    \label{fig:7}
\end{figure}
In an attempt to discern the subsequent dynamo effects induced by different subgrid physics, we have performed several simulations without feedback and star formation. These simulations still include radiative cooling and a Jeans floor. This removes one of the main contributors (feedback) of turbulence and vertical motions from the simulation, leaving the gravitational collapse and shear as the determinant factors. The underlying kinematics will thus be highly dependent on the cooling and the given Jeans floor (setting the minimum collapse-length). We perform two sets of simulations using the initial vertical and toroidal magnetic fields of \eq 8 and 9, and with varying Jeans floor ($N_J=4,8,10,16$) to investigate how the the magnetic field amplifies in different conditions. The simulations are run up to varying times depending on saturation conditions of the magnetic field. The results of the simulations are shown in \Fig \ref{fig:1} to \ref{fig:7}.
\\ \\
In \Fig \ref{fig:1} we can see the state of the galactic disk of our different runs at around the same time. We can see that we get very different amplification and behavior when changing the Jeans floor. Lower Jeans floor allows for more collapse of the gas and stronger spiral arm dynamics within the disk, which seems to greatly increase the amplification of the magnetic field within the disk. There is, however, not a linear dependence on the magnetic field amplification and lower Jeans floor. The $N_J=4$ case initially experiences a quick amplification of the magnetic field within the collapsing spiral arms. However, this amplification is eventually damped as the spiral arms fragment and lose interconnectivity. If we compare this to the $N_J=8$ case we can see that while the spiral arms in this case also fragment, there exists more elongated spiral arms and higher connectivity between the fragments. Amplification in this case occurs rapidly, reaching a saturated state after around $500$ Myr (\Fig \ref{fig:2}). The radial profile of the magnetic, turbulent and thermal energy densities (\Fig \ref{fig:3}) reveals that the magnetic field reaches equipartition in the centre of the disk, which strength then tapers off as we go to larger radius. Looking at the evolution of the averaged $B_{rms}$ within the disk, we can see that it is quickly amplified to about $10\mu G$, where it eventually saturates. From \fig \ref{fig:2} we can see that during the amplification stage the average $\alpha_{MW}$ peaks within the disk. During the evolution of the galaxy, we observe plenty of field reversals within and around the spiral arms, where the main amplification takes place, indicating that we have an active dynamo cycle acting in this region. These spiral arms strongly interact with each other as they move radially through the disk. This can lead to strong magnetic field amplification, as the spiral arms gets entangled. In addition, the spiral arms oscillate in the vertical direction (around 100 pc), which increases the entanglement of the spiral arms as they interact, which can lead to further magnetic field amplification.
\\ \\
As we increase the Jeans floor even further ($N_J=10$ and $N_J=16$), we can see that we have less collapse, and at the time of \fig \ref{fig:1} there is no significant fragmentation of the disk. In $N_J=16$ the spiral compression is highly reduced, which dampens the magnetic field amplification. The amplification from shear and small turbulent motions in the disk solely remain to balance out the dissipation/diffusion of the magnetic field. In the high Jeans floor cases an apparent difference between the initial magnetic field orientation arises. From \fig \ref{fig:1}, we can see that in the case of a toroidal field the centre region becomes highly damped. Furthermore, in the case of $N_J=16$ there is dissipation throughout the whole disk. The dampening of the central region in the toroidal cases can also be seen at early times for $N_J=8$ before significant amplification has occurred. We believe that the main reason for this is that particles which smoothing kernel crosses the central axis of the disk will "see" a very discontinuous magnetic field in the case of an initial toroidal field and the artificial resistivity will attempt to smooth it out leading to high dissipation of the field in this region. This will not be the case with an initial vertical field as it will vary smoothly across the axis. In addition, for both cases the central region will have a significant reduced ability to amplify the magnetic fields compared to the outer regions due to a lower shearing parameter (see \fig \ref{fig:4}). As we mentioned previously, amplification in these high Jeans floor cases will be driven strongly by shear and the vertical motions of the turbulence. Apart from the numerical resistivity, there is additionally also turbulent diffusion and advection of the mean magnetic field that can affect the amplification. Divergence cleaning might at first also seem like a probable cause for the dampening within the centre region, however, we have tested without divergence cleaning and the dampening still remains.
\\ \\
Looking at the polarity of the $N_J=16$, $B_{0,z}$ disk in \fig \ref{fig:5}, we can see that we have magnetic field reversals in both the developed radial and toroidal field structure. These field reversals primarily form in the inter-arm regions, and have been shown to be associated with the velocity changes across the spiral shocks that form within the disk \citep{2016MNRAS.461.4482D}. The vertical field remains similar to the initial field structure, with the magnetic field strength continuing to be highly correlated to the density.
\\ \\
An interesting feature can be seen in the case of $N_J=10$ for an initial vertical field, at around $x=-6$, $y=-6$ in \fig \ref{fig:1}, where a fragment has started to develop. Running this simulation for longer, we can see from the time lapse in \fig \ref{fig:6}, that this fragment together with its connecting filaments causes an instability to occur in the disk, leading to a subsequently rapid amplification of the magnetic field, similar to what was seen in the $N_J=8$ case. Looking at the magnetic field generated around the fragment in \fig \ref{fig:7}, we can see that it is highly entwined with the connecting high dense filamentary structure, stretching the field towards the radial direction. In the low density region in front of the fragment we can see that the field reverses its direction. As time goes on, the fragment and filaments move radially inward as can be seen in \fig \ref{fig:6}. In addition, the fragment oscillates in the vertical direction around the central plane. Strong magnetic fields can be seen to be generated in the wake of the fragment as it sweeps up gas.
\\ \\
Strong candidates of the dynamo action induced in the spiral arms in these simulations is an $\alpha\Omega$ type dynamo, where radial fields are induced by either large-scale motions as in the GI-dynamo or turbulent motions as in the classic $\alpha\Omega$ dynamo. In the GI-dynamo we expect large-scale vertical rolls to be generated above the disk. While there is significant vorticity within the velocity field above the disk, clear vertical rolls cannot be seen and the velocity field looks highly turbulent.  The strongest amplification can be seen in the filamentary structures that move inward in the disk and has magnetic fields that is strongly dragged along the radial direction. This can partly be understood by considering the terms governing the generation of radial fields within mean-field dynamo theory. The electromotive force (EMF) involved in generating the radial fields has two main components, the part that originates from vertical motions ($\overline{u_z b_r}$) and the part that originates from radial motions ($\overline{u_r b_z}$). Radial fields generated through the pinching of the field lines within the filaments are lifted by vertical motions (either turbulent or large-scale) that redistributes the radial fields in the vertical direction, leading to a segregation of the magnetic flux, giving an opposite positive/negative mean-field within the filament and in the corona. These vertical motions can in addition induce vertical magnetic fields $b_z$ that together with radial motions $u_r$ can stretch and fold the field lines in the radial direction. The collective effect of these two components will depend on how net-correlated the magnetic field and velocities are. The effect seen here replicates many of the features seen by the GI-dynamo as described by \cite{2019MNRAS.482.3989R}, where radial flux redistribution within the spiral arm was found to generate opposite mean radial magnetic fields within the spiral arm and its surroundings (corona and interarm region). This is similar to the phenomena that we witness in \fig \ref{fig:7}, where the magnetic field can be seen to be highly connected to the filamentary structure, with opposite/unstructured magnetic fields occurring in the low density region around it. We can in addition see that fluctuating vertical fields are increased within the filament region. \\ \\ \\ \\ The small vertical bulk motions of the filament and fragment might add extra complexity to the dynamo processes as the vertical density structure around it will become more asymmetric as it moves away from the central plane. In the case of $N_J=8$ there is also significant interaction between the spiral arms that likely boost the amplification of the dynamo. A full mean-field analysis is required to separate the effective scales and the contribution of each component. This would allow one to more easily distinguish between the GI-dynamo and the classic $\alpha\Omega$ dynamo in this case. This is beyond the scope of this paper and we leave it to be explored in future work.
\subsection{Simulations with Feedback}
\subsubsection{The early amplification phase and its dependence on resolution}
\label{sec:resfb}
\begin{figure*}[!ht]
    \centering
    \includegraphics[width=\hsize]{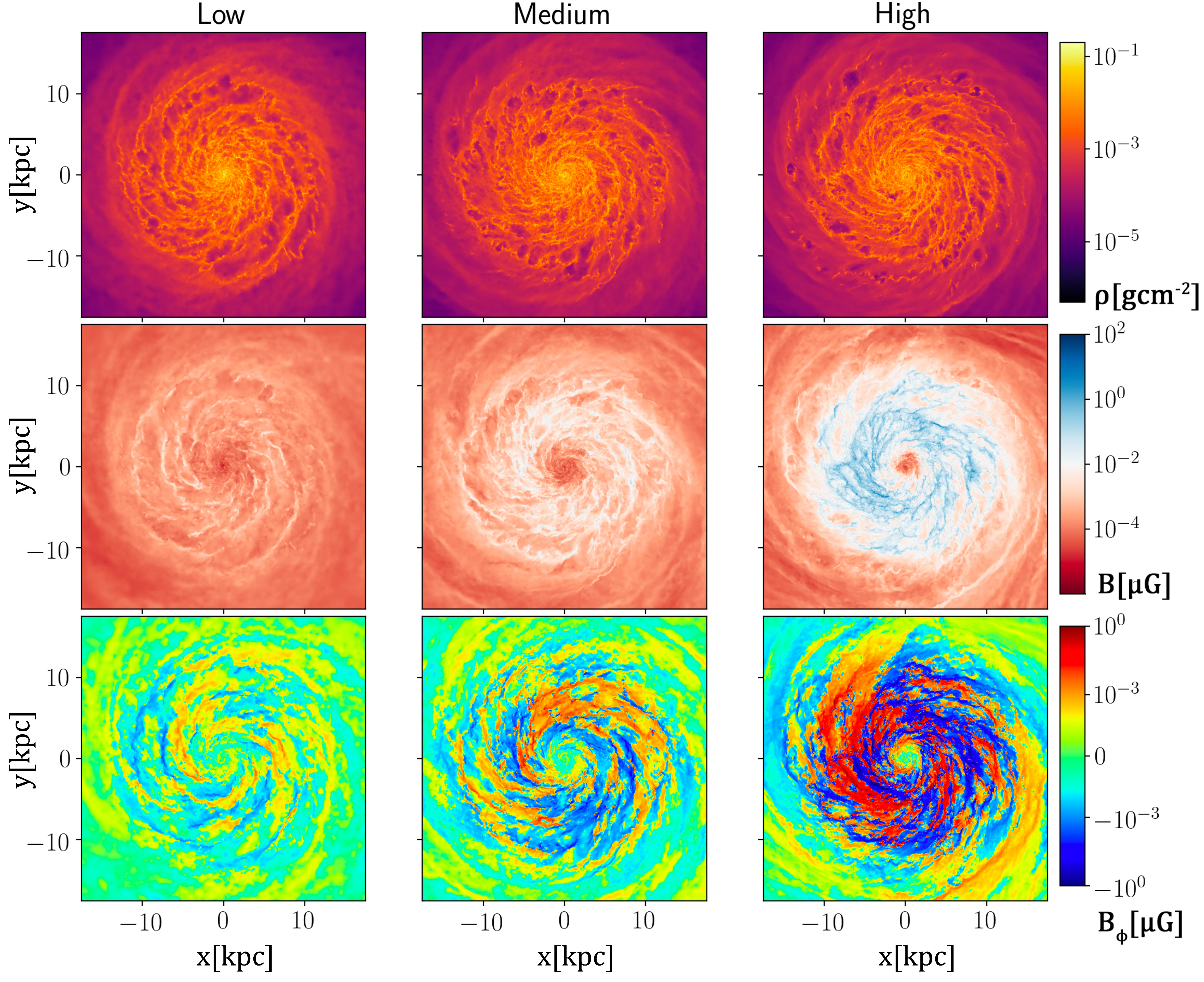}
    \caption{Face-on rendering of the galactic disk after $t=250$Myr. Comparing the effect of resolution on the early evolution of the galaxy. Left to right goes from low to high resolution  Top panel show density, middle panel magnetic field strength and bottom panel the toroidal magnetic field. We can see that the scale of the developed mean-fields are about the same but increases in strength with resolution.}
   \label{fig:8}
\end{figure*}
\begin{figure*}[!ht]
    \centering
    \includegraphics[width=14cm]{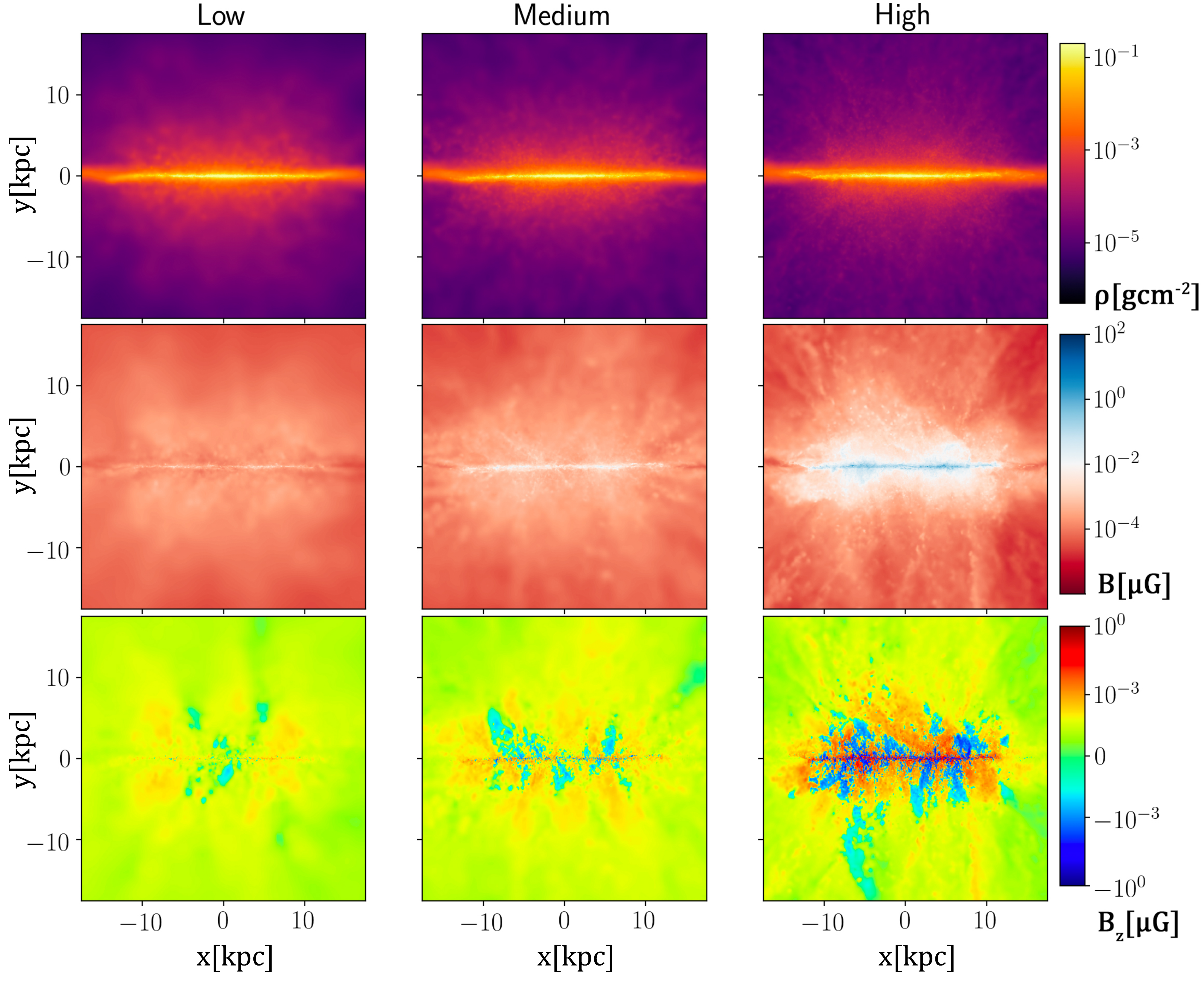}
    \caption{Side-on rendering of the galactic disk after $t=250$Myr. Comparing the effect of resolution on the early evolution of the galaxy. Left to right goes from low to high resolution  Top panel show density, middle panel magnetic field strength and bottom panel the toroidal magnetic field. The increase in resolution show additional the CGM. Feedback leads to the formation of an interesting vertical structure of the magnetic field, where near the central disk, we can see plenty of reversals and intricate behaviors in the magnetic field. The complexity of the structure around the central disk increases as we increase the resolution. The outer regions can be seen to be dominated by the blowout of the initial magnetic flux.}
    \label{fig:9}
\end{figure*}
\begin{figure}[!ht]
    \centering
    \includegraphics[width=\hsize]{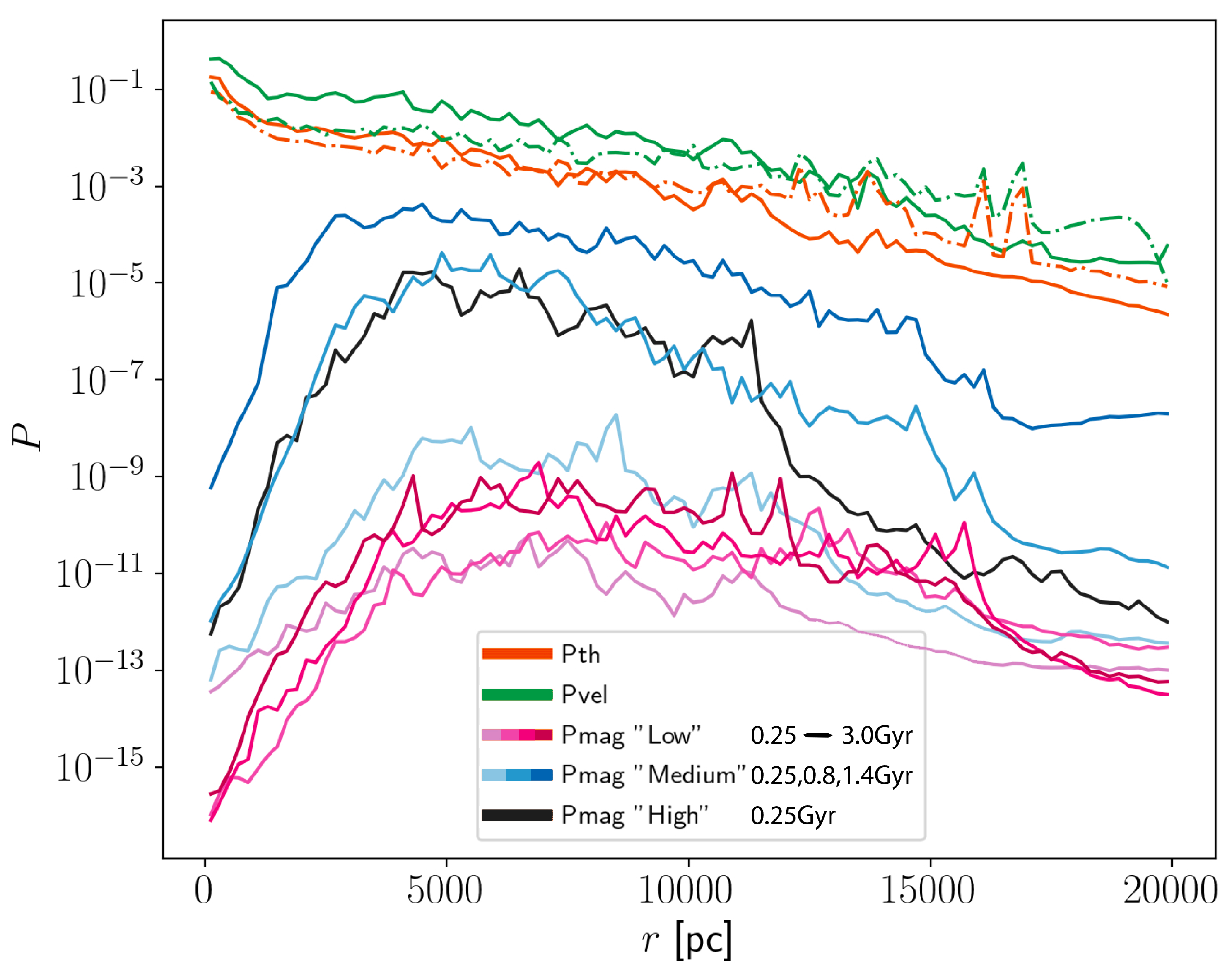}
    \caption{Radial profiles of the magnetic, turbulent and thermal pressures for the resolution study at a few different times (later times represented by given color palette getting darker). For the low resolution simulations we do not see any significant amplification as time goes by. While for the medium resolution we reach a saturated state after about 1.4Gyr. The high resolution have reached a similar magnetic field strength to the medium resolution at 800Myr after just 250Myr.}
    \label{fig:10}
\end{figure}
In the following sections we look at the effect of adding star formation and feedback to our simulations. In this section we investigate the early amplification phase of the magnetic field and its dependence on resolution. To reduce the computational cost of our highest resolution simulation, $N_{FB}=1$ is used for the simulations in this section. Apart from this, the initial configuration as outlined in the beginning of \sect \ref{sec:results} is used. We run three simulations at different resolutions (\tabl\ref{table:1}), which we will refer to as the "Low", "Medium" and "High" simulations. The Jeans floor adjustment is taken into account to resolve the same collapse-scale across all resolutions. This correspond to resolving the Jeans length by $N_J=4,8,16$ for the "Low", "Medium" and "High" simulations, respectively. The "Low" and "Medium" simulations are run for around $2$ Gyr, while the "High" simulation is only run for $250$ Myr due to being computationally demanding. The results of the simulations are shown in \Fig \ref{fig:8} to \ref{fig:10}.
\\ \\
In \fig \ref{fig:8} and \ref{fig:9}, we can see the state of the galactic disk of our resolution study after around $250$ Myr. It is clear that we have a strong resolution dependence on the amplification of the magnetic field. Due to the cold initial conditions, there is a starburst in the beginning of the simulation. This leads to strong initial outflows that advects the magnetic field outwards and a temporary decrease in the magnetic field within the disk. But the magnetic fields are quickly strengthen by the dynamo processes active in the disk. The amplification of the magnetic field can be seen to mainly occur in the spiral arm region of the disk, whereas in the centre of the disk there is no / much less amplification. This is highlighted in \fig \ref{fig:10}, which shows the radial profile of the turbulent, thermal and magnetic pressure. It is clear that the shape of the magnetic pressure curve is similar across all the simulation, indicating a similar amplification process across all the simulations. This is further seen in the scale of the toroidal mean-fields generated within the disk (bottom panel in \fig \ref{fig:8}), which polarity remain at a similar scale across the three resolutions. While all three resolutions exhibit a similar mean-field dynamo, there is a clear resolution dependence on the effective growth. 
\\ \\
In \fig \ref{fig:10} we have additionally plotted the radial profiles of the magnetic pressure at later times for the 'Low' and 'Medium' resolution simulations. The 'Medium' resolution subsequently amplifies and reaches a saturated state of around $10\%$ of equipartition with the thermal energy, which occurs at around 1200 Myr. The 'Low' resolution, on the other hand, do not exhibit any significant amplification and roughly keeps the strength of the initial magnetic field (averaged over the disk). At the same time, we can see that the 'High' resolution simulation has already amplified the magnetic field at $t=250$ Myr to similar strengths as the 'Medium' resolution simulation at $t=800$ Myr. While the radial pressure shape of the 'Low', 'Medium' and 'High' is similar, there are some interesting differences. The 'Medium' and 'High simulations that amplify the magnetic field have a peak of around 4-6 kpc, with a fairly constant plasma beta ratio between the thermal and turbulent velocity between 4-15 kpc. For the 'High' resolution simulation we can see that we have a stronger drop of at 12 kpc. This is likely due to the simulation being at a much earlier time ($t=250$ Myr) than the comparative 'Medium' resolution curve ($t=800 Myr$), which indicate that 4-12 kpc is the region with highest amplification, correlating to the region which exhibit most feedback bubbles and spiral arms. As time goes on, the magnetic field can be seen to be amplified in the outer and central regions of the disk, through both the advection and diffusion of the strong field regions and potentially through a slower dynamo amplification process in these regions. Feedback also leads to the formation of an interesting vertical structure of the magnetic field, where near the central disk, we can see plenty of reversals and intricate behaviors in the magnetic field (see bottom row in \fig \ref{fig:9}). Further out from the disk we can see that the magnetic field is mainly dominated by the initial vertical flux that is blown out early on in the simulation. Additional structure can be seen to emerge in this low density region as we increase the resolution.
\\ \\
Due to the formation of large-scale mean-fields, it is likely that we have a mean-field dynamo acting in these simulations. Small-scale dynamo can of-course also be active within these simulation, but would not be able to generate the observed mean-fields. We will discuss the potential and effectiveness of small-scale dynamo in these simulation further in \sect \ref{sec:diffsims}. While both these runs and the no-feedback runs appear to amplify due to a sort of $\alpha\Omega$-dynamo, there are some difference between the effective amplification processes. This can clearly be seen in the developed radial profile of the magnetic pressure in \fig \ref{fig:3} and \fig \ref{fig:10}. In the no-feedback run the magnetic field strength becomes concentrated in the centre of the disk as the fragments move radially inward through the disk. While in the feedback runs the magnetic strength is concentrated outside the central region, keeping a similar ratio between the thermal and magnetic pressure in the range $4-16$ kpc. 
\\ \\
The major difference lies in the expansion of the vertical scale-height of the galaxy as we introduce feedback. This causes an increase in magnetic flux transfer from the central disk region to the CGM and an overall increase in the simulated system volume. This increases the magnetic strength in the CGM, but can do so at the expense of the magnetic field within the central disk. On the other hand, the increased vertical motions induced by feedback act to increase the effectiveness of $\alpha\Omega$ type dynamos. The increase in resolution allows for more small-scale structure in the CGM and seem to correlate to a more effective mean-field dynamo in the disk. It is reasonable that this increase in small-scale structure further increase the effectiveness of the dynamo, as it would lead to more resolved flow structure (for example the vertical rolls in the GI-dynamo). From \fig \ref{fig:4}, we can also see that the superbubble feedback in general increases the turbulence in the disk and leads to a higher mach number within the disk. This can have both a positive and destructive effect on the growth of magnetic fields. Increased turbulence will have a positive effect on both small-scale and mean-field dynamo processes as it is the main driver. For the small-scale dynamo, the increase in turbulence mainly has a positive effect on the growth rates given that the fluid parameters ($Re,Re_{mag},P_m$) are high enough. For the mean-field dynamo, the turbulence act to increase the vertical and radial motions, which benefit the amplification. However, for the mean-field, turbulence can also act in a destructive fashion, by increasing the turbulent diffusion within the disk and increasing the ejection of magnetic flux from the central plane. The dampening of the central region likely occurs due to the destructive effects being dominant in this region. We further discuss the effect of resolution in the following sections but with the addition of changing other parameters.
\subsubsection{Effect of Jeans Floor}
\label{sec:jeanfb}
\begin{figure*}[!ht]
    \centering
    \includegraphics[width=\hsize]{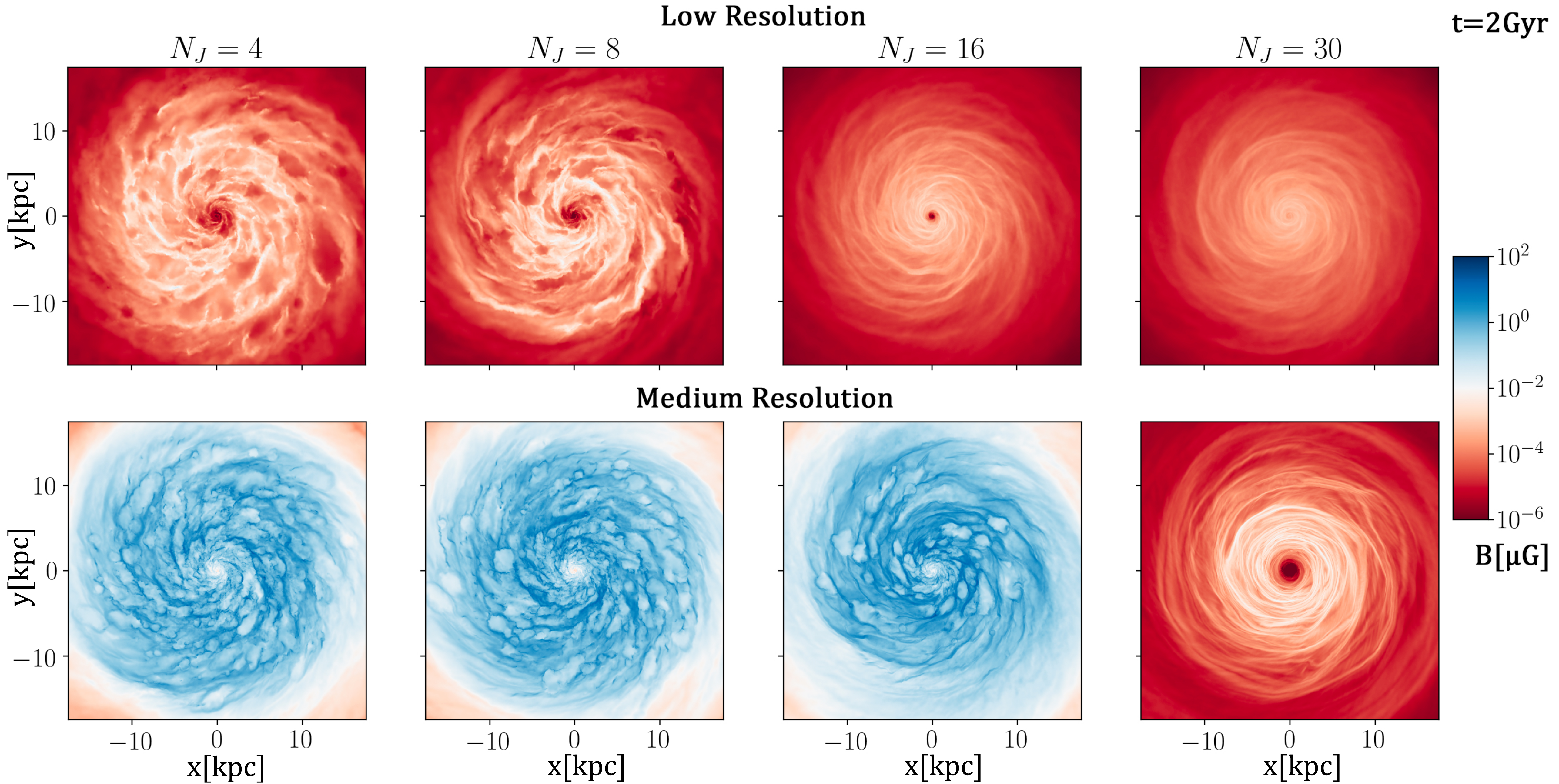}
    \caption{Face-on rendering of the magnetic field strength of galactic disk after $t=2.0$ Gyr. Simulations with feedback, comparing the effect of different Jeans floor on the evolution. Top panel show the Low resolution simulation and bottom the medium resolution. We can see that we get no significant amplification in the low resolution, while strong amplification is seen in the medium resolution. Too high Jeans floor diminishes the dynamo processes as the smallest collapse length becomes too large.}
    \label{fig:11}
\end{figure*}
\begin{figure}
    \centering
    \includegraphics[width=\hsize]{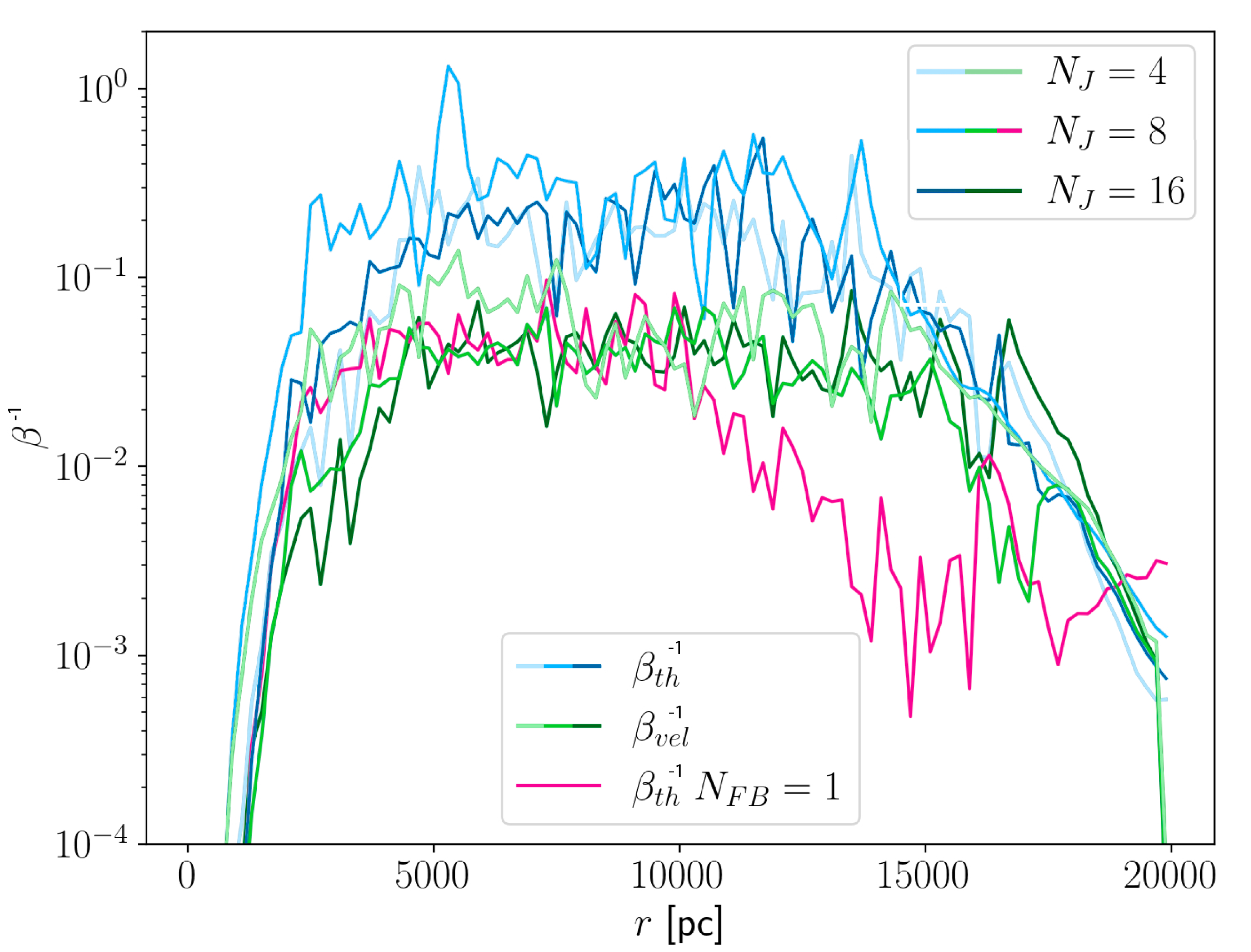}
    \caption{The radial profile of the inverse plasma beta for the saturated medium resolution runs in \fig\ref{fig:11}. We have included lines for both the inverse thermal plasma beta and the inverse turbulent plasma beta. The purple line shows the thermal inverse plasma for the saturated radial profile with $N_{FB}=1$ shown in the previous section. We can see that most of the cases saturate to around $10-30\%$ of the thermal energy density. With $N_{FB}=1$ the saturation is lower in the outer regions.}
    \label{fig:12}
\end{figure}
\begin{figure*}
    \centering
    \includegraphics[width=\hsize]{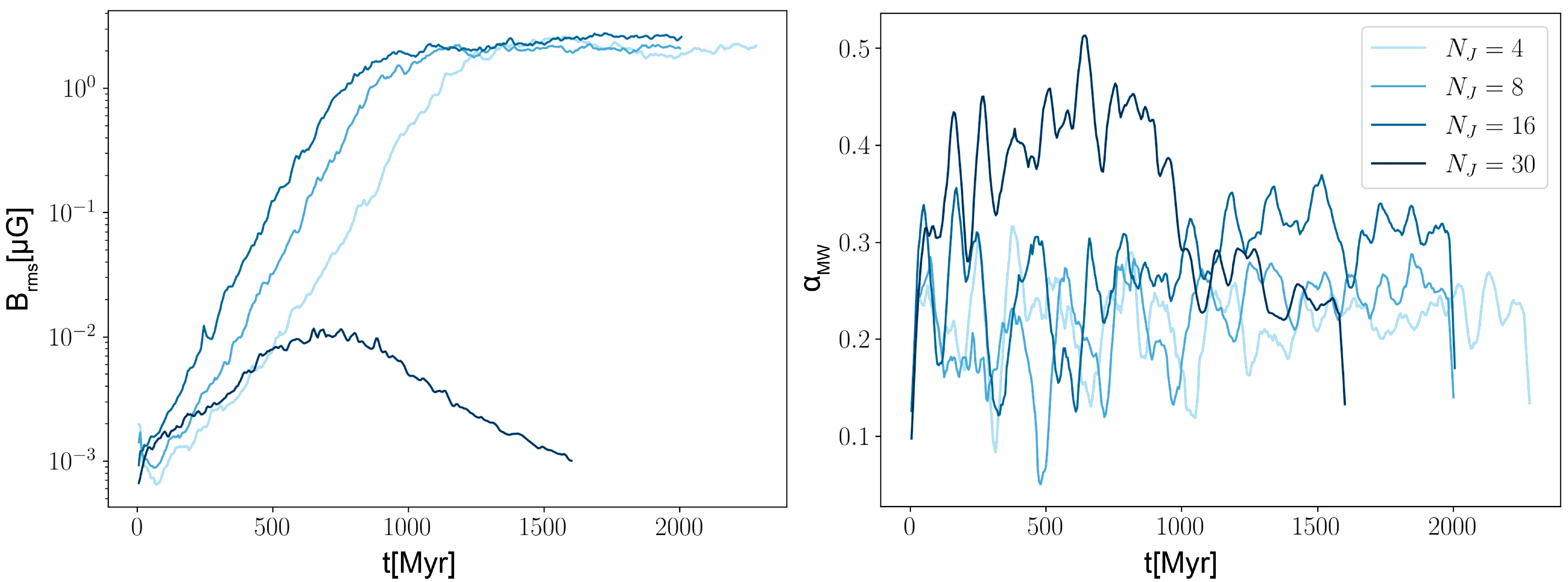}
    \caption{Time plots for the medium resolution feedback runs with varying Jeans floor. The left and right panel shows the evolution of $B_{rms}$ and $\alpha_{MW}$ respectively. Darker colors represents higher Jeans floor. We can see that there is a positive dependence on the Jeans floor for the amplification rate, as long as a "critical" collapse-length is resolved. The normalized Maxwell stress lie around $\alpha_{MW}=0.1$-$0.3$ for all runs. There is additional stress within the $N_J=30$ simulation during the early amplification phase with around $\alpha_{MW}=0.4$.}
    \label{fig:13}
\end{figure*}
\begin{figure}
    \centering
    \includegraphics[width=\hsize]{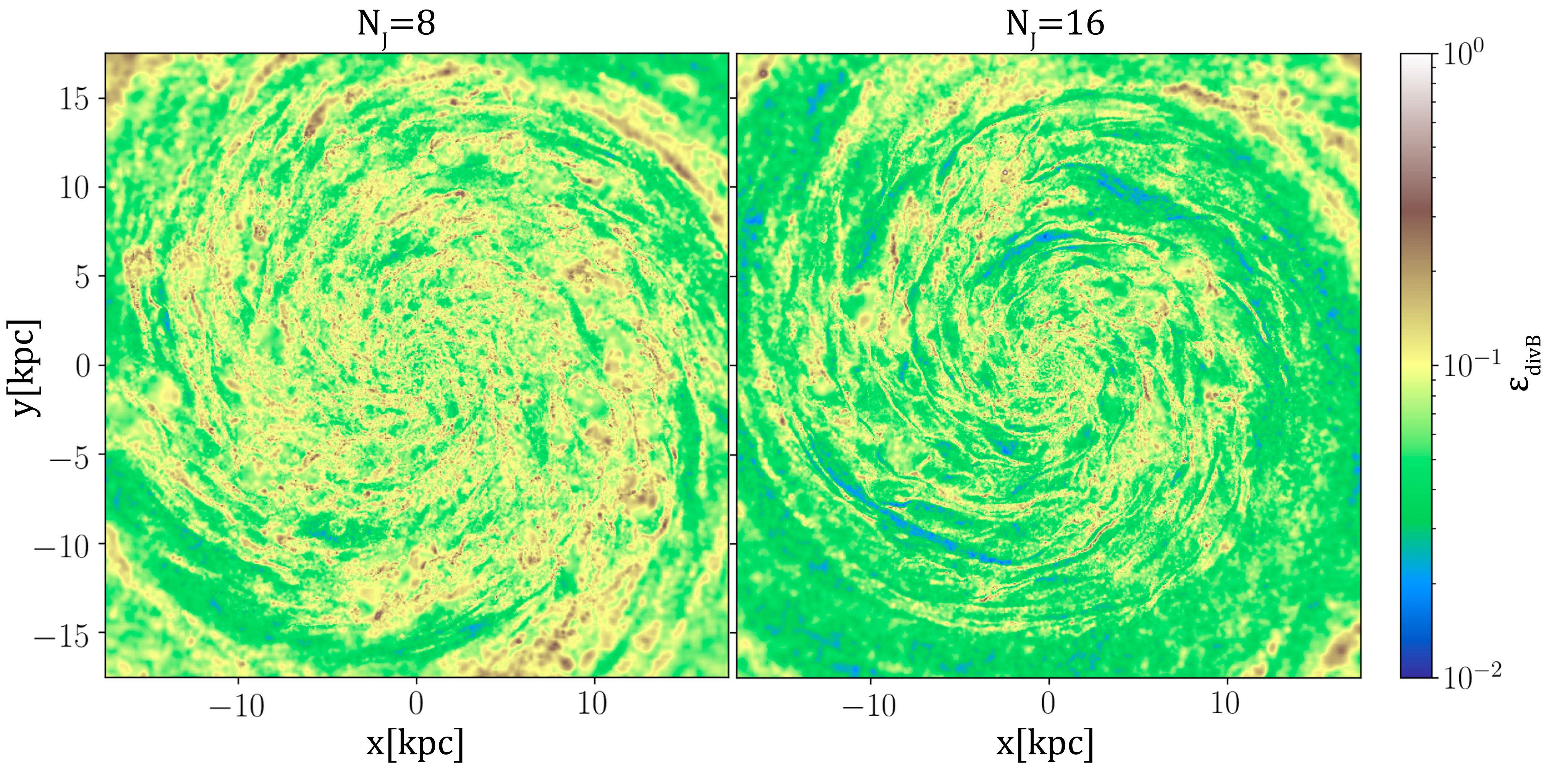}
    \caption{Face-on rendering of the divergence error of the magnetic field in the galactic disk after $t=2 Gyr$. We can clearly see that there is a reduction in divergence error throughout the disk for higher Jeans floor. The mean divergence error in the $N_J=8$ is around $\epsilon_{divB}=10^{-1}$ and in the $N_J=16$ around $\epsilon_{divB}=6 \cdot 10^{-2}$.}
    \label{fig:13divb}
\end{figure}
This section aims to investigate the effect of altering the numerical Jeans floor of our feedback simulations. This is similar to the investigation done in \sect \ref{sec:nofb} for our no-feedback runs. We vary the Jeans floor between ($N_J = 4, 8, 16, 30$) and run two sets of simulations at 'Low' and 'Medium' resolution (see \tabl\ref{table:1}). For everything else the default initial values are used (\sect \ref{sec:results}) and the simulations are run to around $t=2 Gyr$. The results of the simulations are shown in \Fig \ref{fig:11} to \ref{fig:13divb}.
\\ \\
In \Fig \ref{fig:11} we can see a rendering of the magnetic field strength in the galactic disk between all our runs at $t=2 Gyr$. The most stark difference seen is the effect of resolution, where $1-30 \mu G$ develop in the 'Medium' resolution while barely any amplification occurs in the 'Low' resolution. This is similar to what we observed in the previous section when we looked at resolution dependence. Similar to the no-feedback runs, we can see that the galactic disk experience much less dynamics when the collapse-length is too large, above $N_J=16$ for the 'Low' resolution and above $N_J=30$ for the 'Medium' resolution. Below these $N_J$ values we can see that the magnetic field saturates to a similar value across the simulations independent of the Jeans floor. This is further highlighted in \fig \ref{fig:12} where we have plotted the radial profile of the inverse plasma beta for the saturated 'Medium' simulations ($N_J=4,8,16$). In between $4-15$ kpc, we can see that all the runs saturate at around $10-30\%$ of the thermal pressure, while reaching only around $5\%$ of the turbulent pressure. We can see that the $N_J=8$ reaches higher saturation in the inner region $2-4$ kpc compared to the other cases. It is interesting that if one compares these runs to the $N_{FB}=1$ 'Medium' run from the previous section, the increase in the number of feedback particles in these simulations produces a significant increase in the saturation strength beyond 10 kpc (as seen in \fig \ref{fig:12}). However, this is mainly due to the $N_{FB}=1$, having hotter winds and resulting higher thermal energy in the outskirts. The main dependence on the Jeans floor seem to arise when we look at the time evolution between these different runs. As can be seen from \fig \ref{fig:13}, there is a positive correlation between the growth rate of the magnetic field and the Jeans floor, given that the collapse length is sufficiently small. Saturation is achieved at around $t=1000$ Myr for $N_J=16$, $t=1100$ Myr for $N_J=8$ and $t=1300$ Myr for $N_J=4$. From this it also becomes apparent that the $N_J=30$ run experiences a growth phase in the magnetic field to about $t=800$ Myr, while then starting to slowly decay. Due to the initial cold state, there is some feedback occurring early in its evolution which increases the dynamics of the galaxy and induces magnetic amplification. However, when the star formation slows down and the galaxy calms down, the magnetic field starts to slowly decay. Compared to the no-feedback runs, we can see that the normalized Maxwell stress remains around $\alpha_{MW}\approx 0.2$ for $N_J=4,8,16$ and there is no significant increase during the growth phase (\fig \ref{fig:13}). For $N_J=30$ there is however, an increase in normalized stress during the growth phase $\alpha_{MW}\approx 0.4$, which reduces to around $\alpha_{MW}\approx 0.2$ as it starts to decay. Another effect that we can see from these simulations are that the divergence error becomes smaller for larger Jeans floor. In \fig \ref{fig:13divb}, we can clearly see that the divergence error within the disk is reduced as we increase the Jeans floor from $N_J=8$ to $N_J=16$. This is likely due to the fact that we have more smooth collapsed structures within the disk, which has better resolved gradients that in turn causes less divergence errors to be produced. The divergence errors are comparable or better than previous Lagrangian codes, which usually lie in the range $\epsilon_{div,B}=10^{-1}\leftrightarrow10^1$ \citep{2009MNRAS.397..733K,2013MNRAS.432..176P,2016MNRAS.461.4482D,2019MNRAS.483.1008S}.
\\
\subsubsection{Effect of Feedback}
\label{sec:fbsims}
\begin{figure}[!ht]
    \centering
    \includegraphics[width=\hsize]{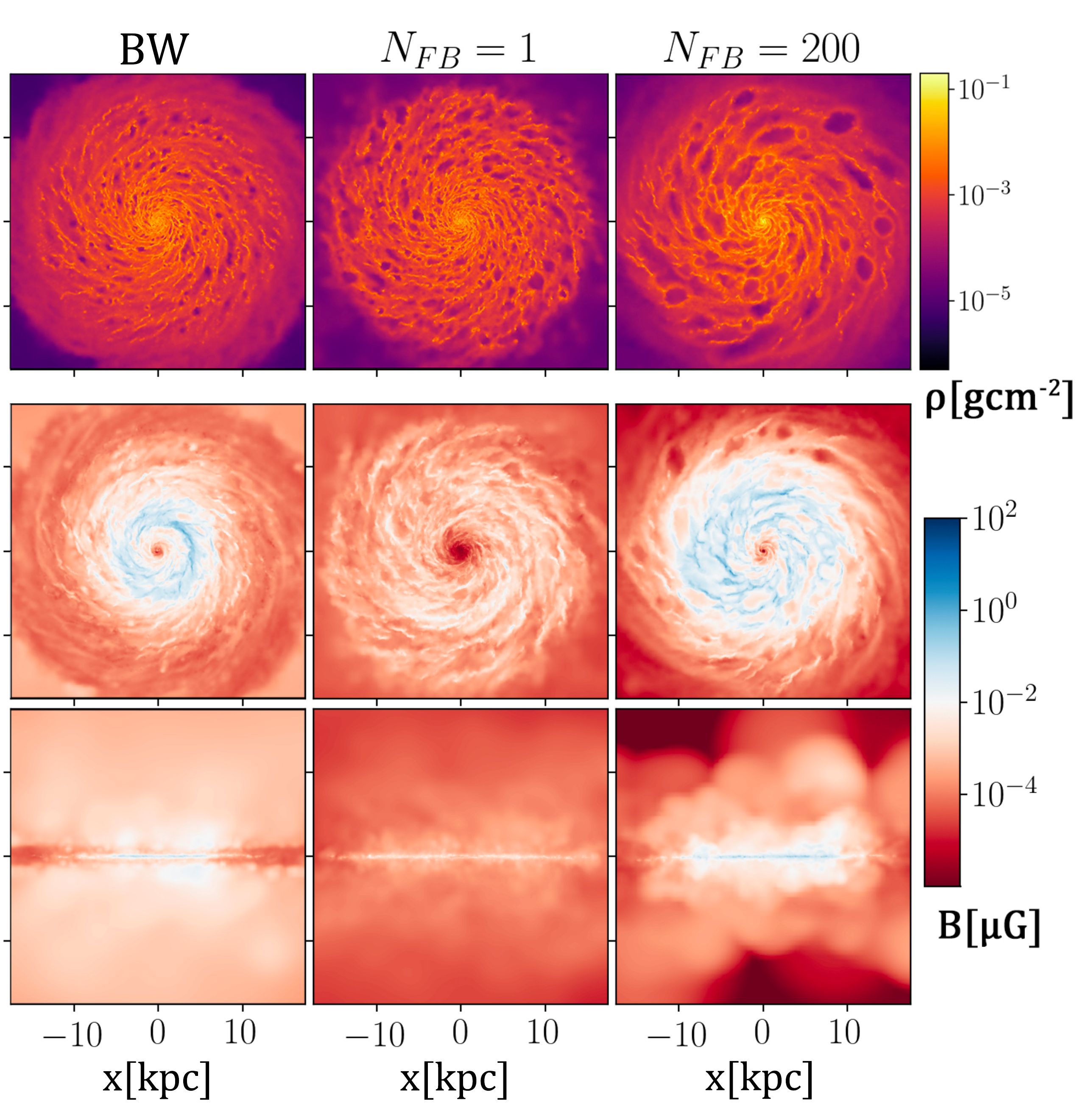}
    \caption{Rendering of density, magnetic field strength of the 'Low' resolution simulations with varying feedback scheme at $t=2$ Gyr. Strong field growth can be seen for both the blastwave model and the higher injection length with the superbubble model. Both the blastwave model and the low injection length leads to small feedback bubbles which pushes material far away from the disk. The mass loading in the superbubble model ($N_{FB}=1$) is however much larger as the blastwave model mainly generate hot low density outflows, this means that there is less transfer of magnetic fields from the central region in the blastwave model. Increasing the injection length leads to larger scale feedback bubbles but a more concentrated vertical structure, which in turn lead to more amplification of the magnetic field within the central disk.}
    \label{fig:14}
\end{figure}
\begin{figure*}[!ht]
    \centering
    \includegraphics[width=\hsize]{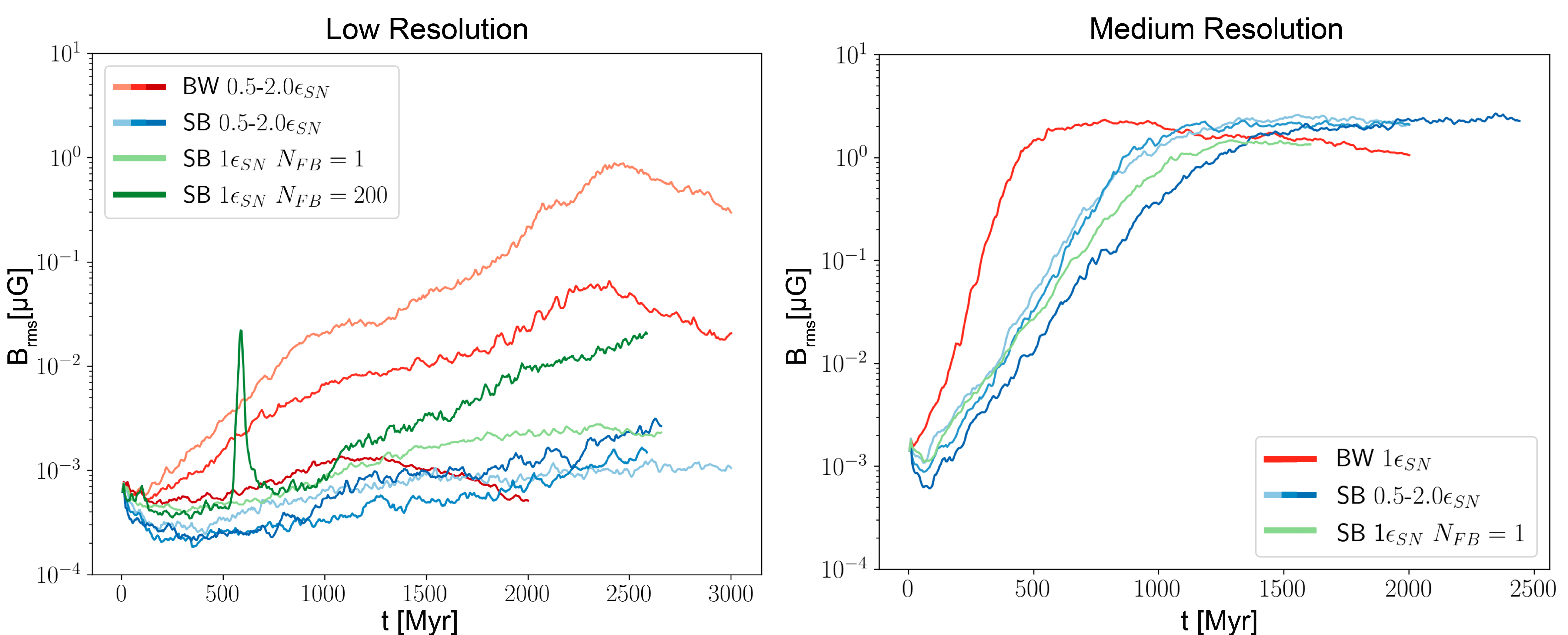}
    \caption{Time evolution of $B_{rms}$ within the central disk for different feedback schemes. Left panel show the 'Low' resolution runs and right panel show the 'Medium' resolution runs. Darker colors in blue and red indicate stronger feedback strength and the green lines represent different injection length from the default $N_{FB}=64$. Its clear from both resolutions that the blastwave scheme produce faster amplification than the superbubble scheme. Higher energy injection does in general lead to slower growth of the magnetic field within the central disk. Increasing the injection length of the superbubble scheme does seem to generally have a positive effect on the magnetic field amplification. The bump seen in the $N_{FB}=200$ evolution is due to rapid amplification of the magnetic field during the merger of two fragments within the central region, which afterwards is quickly dispersed outward from the central disk. 
    }
    \label{fig:15}
\end{figure*}
\begin{figure*}[!ht]
    \centering
    \includegraphics[width=\hsize]{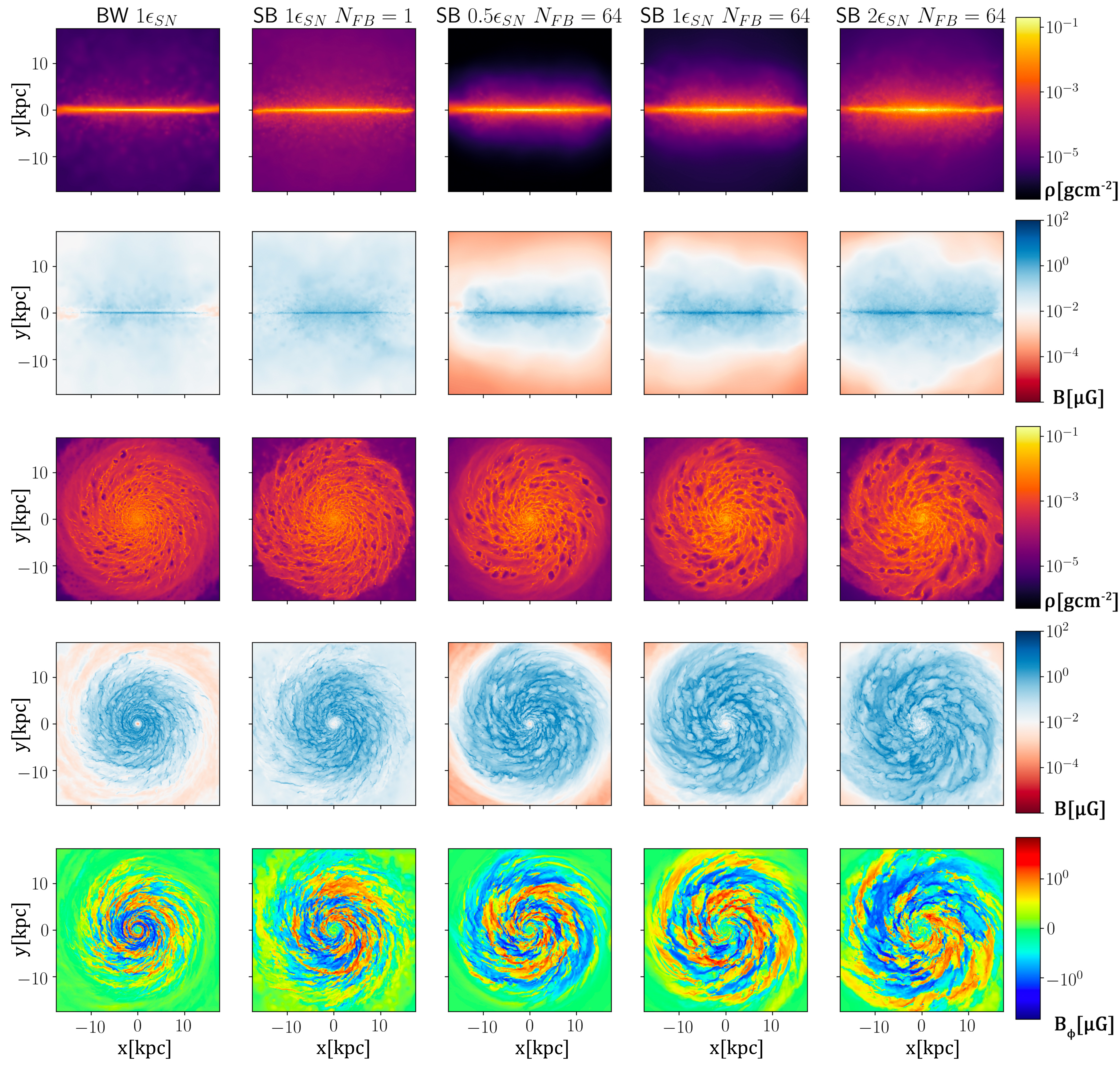}
    \caption{Rendering of density, magnetic field strength and toroidal magnetic field of the medium resolution simulations with varying feedback schemes. The blastwave model generates hotter winds and less mass loading than the superbubble model, leading to less magnetic field advection from the central plane of the disk. This generates a more radial compact structure for the magnetic field. Increasing the injection length in superbubble makes the vertical density structure more compact, leading to more resolved vertical velocity structures. In addition, we see a clear increase in the scale of the toroidal mean-field with the feedback strength as the galaxy extends both radially and vertically.}
    \label{fig:16}
\end{figure*}
\begin{figure}[!ht]
    \centering
    \includegraphics[width=\hsize]{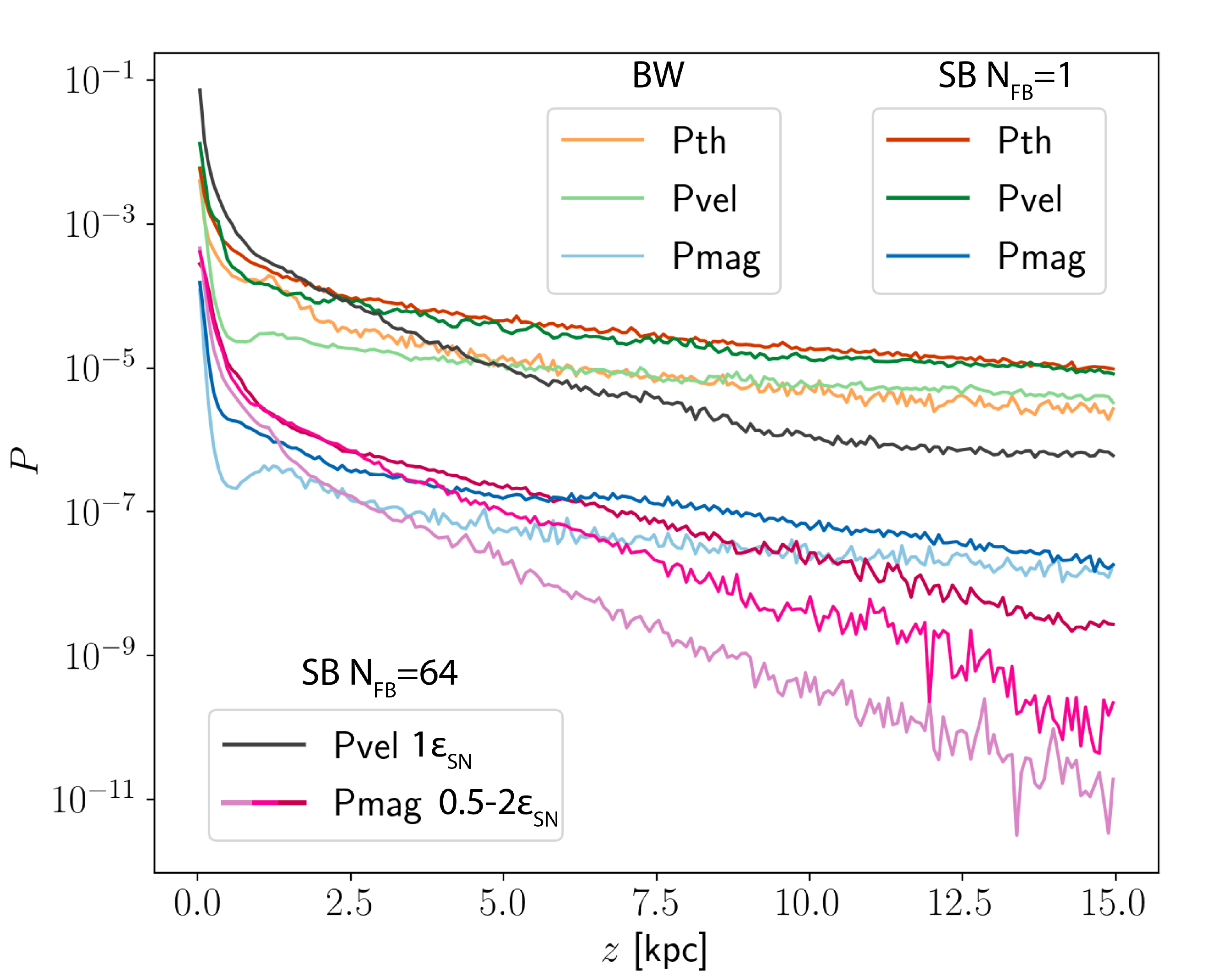}
    \caption{Plot of the vertical structure of the kinetic, magnetic and thermal energy density for our medium resolution runs with varying feedback scheme around $2 Gyr$. The magnetic energy density can be seen to follow the kinetic energy quite closely within the CGM. With stronger magnetic fields being present further out in the strong feedback simulations.}
    \label{fig:17}
\end{figure}
\begin{figure}[!ht]
    \centering
    \includegraphics[width=\hsize]{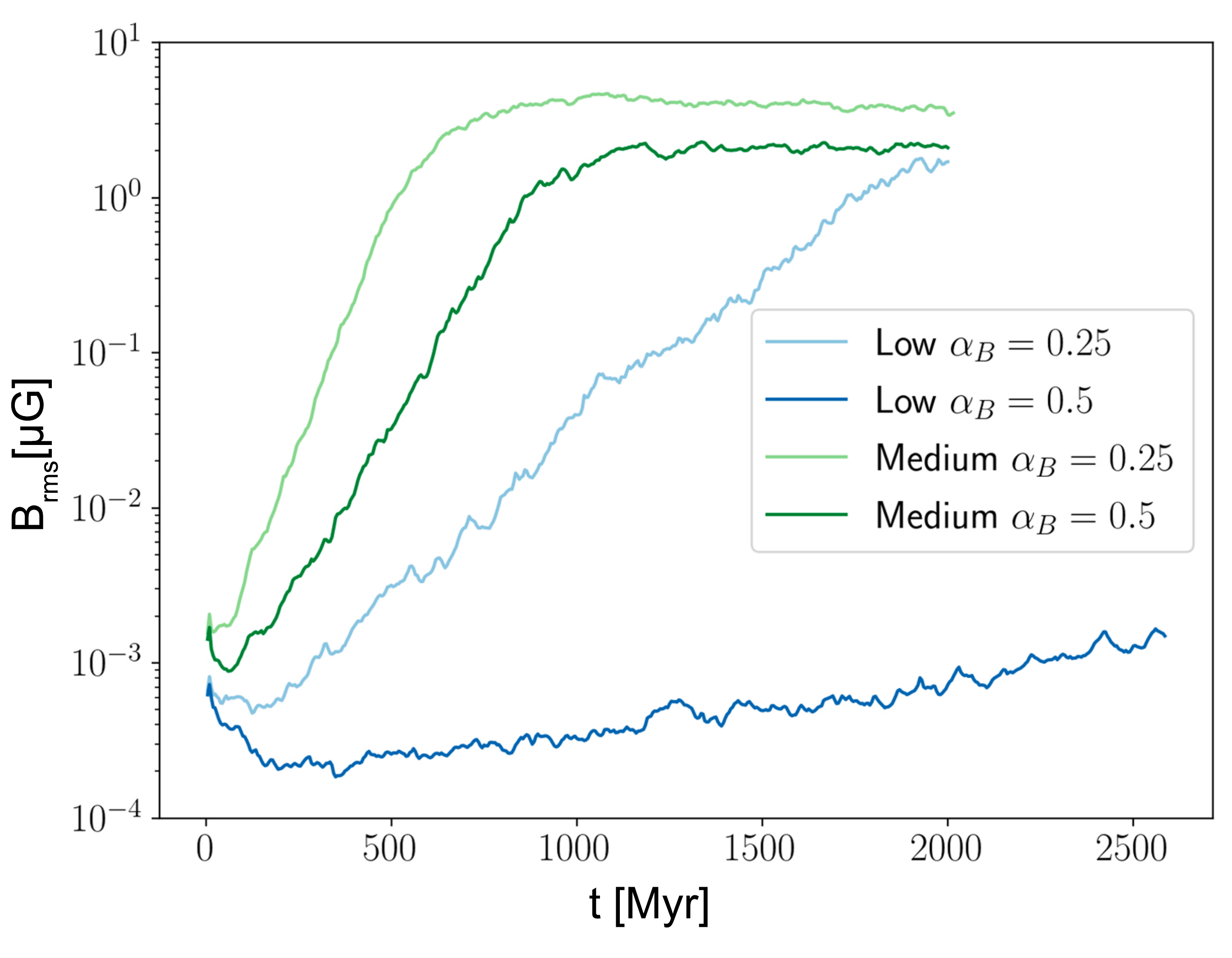}
    \caption{Time evolution of $B_{rms}$ within the central disk for different numerical diffusion and resolutions. We can see that lowering the numerical diffusion has a clear positive effect on the growth rate and saturation level. Even the 'Low' resolution reaches saturation in the case of $\alpha_B=0.25$, after around $1.9Gyr$.}
    \label{fig:18}
\end{figure}
From the previous sections, we have seen that the inclusion of feedback significantly alters the galactic dynamo. Feedback boosts dynamo action in the spiral arms through the injection of vertical fountain motions, but at the same time it can lead to a decrease in magnetic field strength through the dissipation and advection of magnetic flux from the central region. In this section we look at the difference between the blastwave (BW) and the superbubble (SB) feedback models. We also look at the effect of varying the feedback strength ($\epsilon_{FB}=0.5\epsilon_{SN},1\epsilon_{SN},2\epsilon_{SN}$), and for the SB model, we vary the number of neighbours we inject the feedback into ($N_{FB}=1,64,200$). The injection of feedback into more particles than 1 or a few gas particles is not very physically motivated as the ejecta mass is quite low and the two-phase model have been shown to correctly capture the expected mass of the bubble \citep{2014MNRAS.442.3013K}. However, increasing the injection length of the SB model will change the resulting turbulence within the disk, which is an interesting parameter to investigate in terms of the galactic dynamo. We run both 'Low' and a 'Medium' resolution simulations. The simulations are run to around $2$ Gyr and the results are shown in \fig \ref{fig:14} to \ref{fig:16}. 
\begin{figure}[!ht]
    \centering
    \includegraphics[width=0.9\hsize]{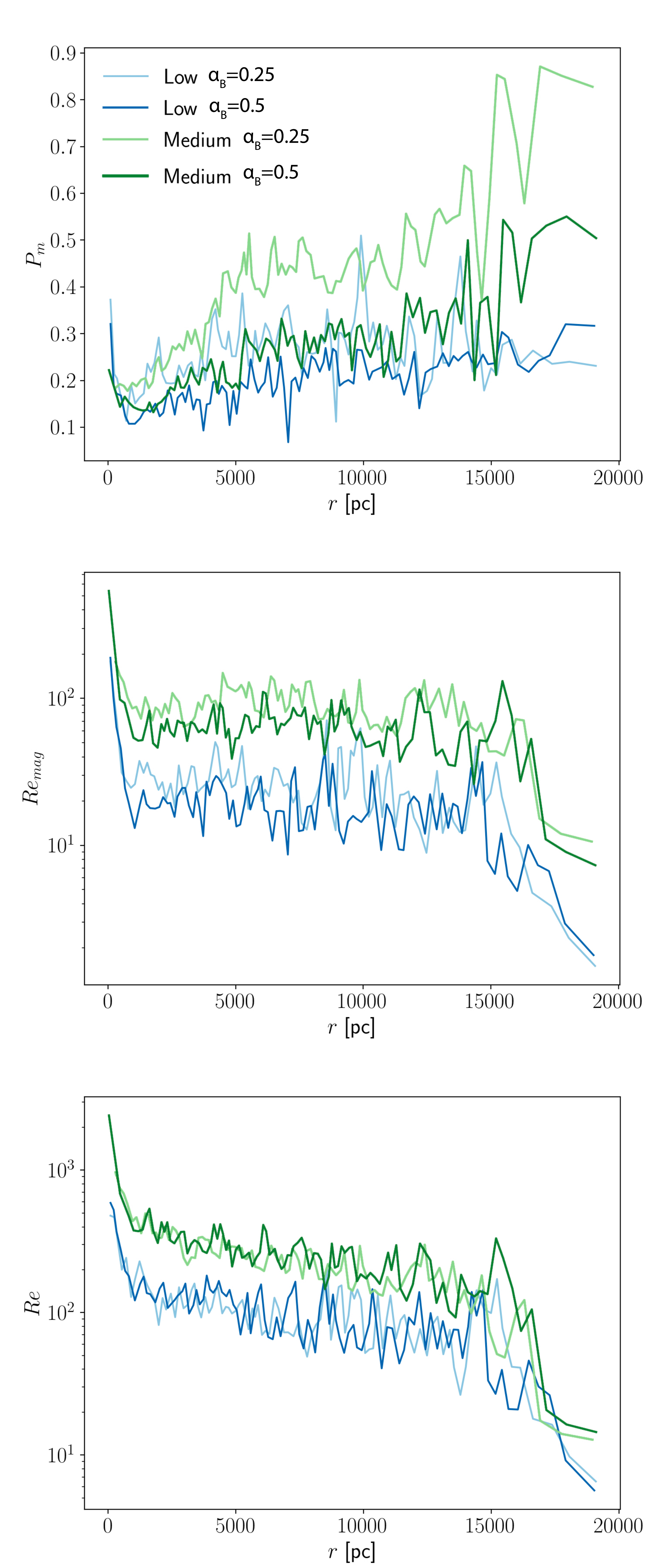}
    \caption{The density averaged radial profile of the fluid parameters for different resolutions and numerical diffusion at $t=2Gyr$. From top to bottom we have: the Prandtl number, the magnetic Reynolds number and the regular Reynolds number. The Prandtl number can be see to be below unity across the whole radial extent, with a decreasing trend towards the centre. There is an expected increase due to lowering artificial resistivty and also an increase in $P_m$ due to resolution in the outer regions of the disk ($\ge 4kpc$). Both the magnetic and regular Reynolds number can be seen to increase towards the centre, however the magnetic Reynolds number remains more flat throughout the disk than its hydrodynamic counterpart.}
    \label{fig:19}
\end{figure}
\\ \\
We first take a look at the effect of the two feedback models. In \fig \ref{fig:14}, we can see result from the 'Low' resolution runs. It is clear that we have stronger magnetic field growth within the inner regions of the BW simulation. Looking at the density structure we can see that BW produces smaller feedback bubbles and a more smooth central region than the SB feedback. In the same figure, the effect of feedback injection can be seen, where there is a positive correlation to the amplification with increasing neighbour number. The density structure show more large scale spiral structure with larger feedback bubble appearing. Within the BW model and the $N_{FB}=1$ model the magnetic field is advected outwards to larger radius, while in the $N_{FB}=200$ we see a more concentrated vertical magnetic field structure. Due to the BW model producing stronger magnetic fields in the disk than the $N_{FB}=1$ model, we see stronger magnetic fields being advected to the CGM.
\begin{figure}[!ht]
    \centering
    \includegraphics[width=0.9\hsize]{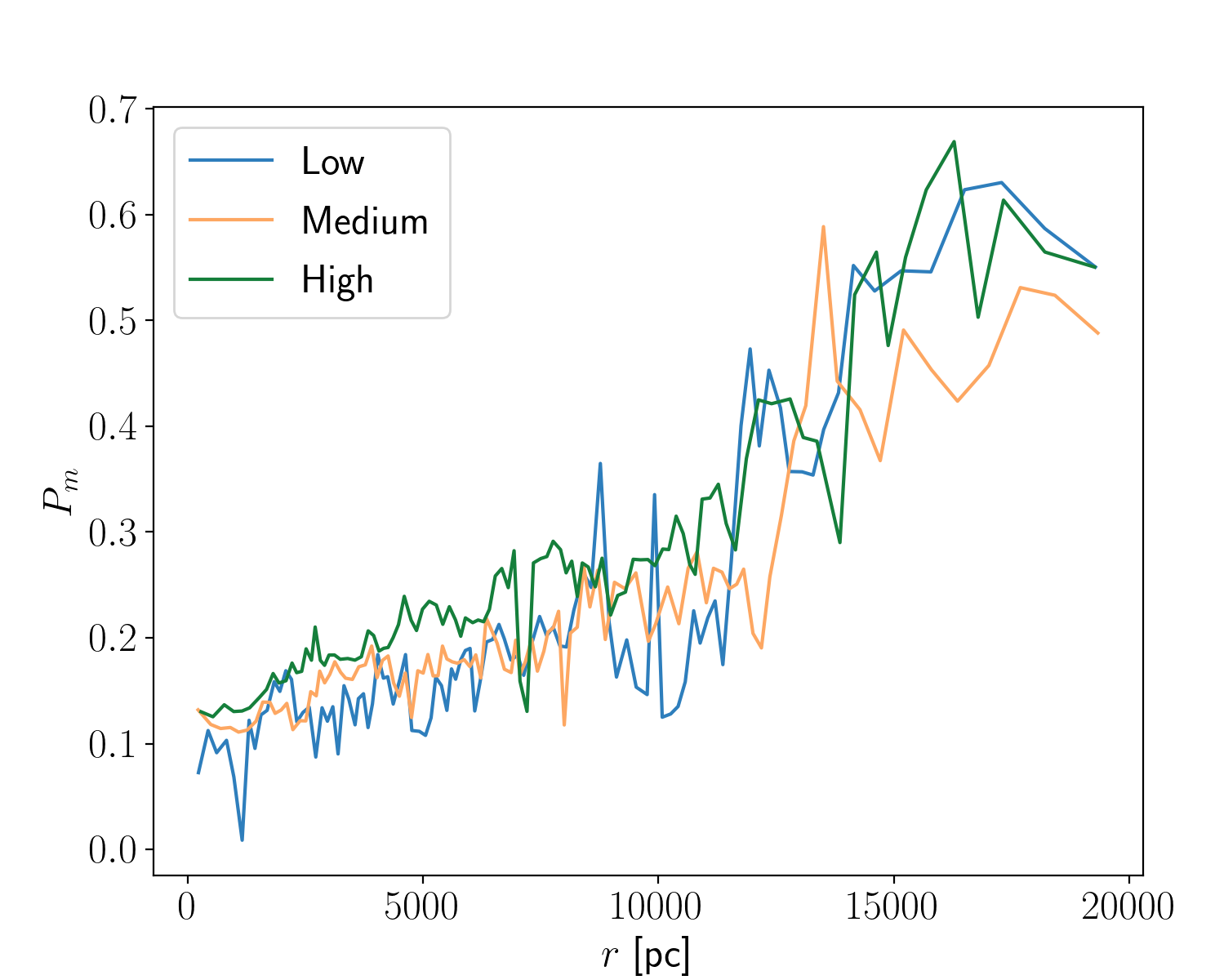}
    \includegraphics[width=0.9\hsize]{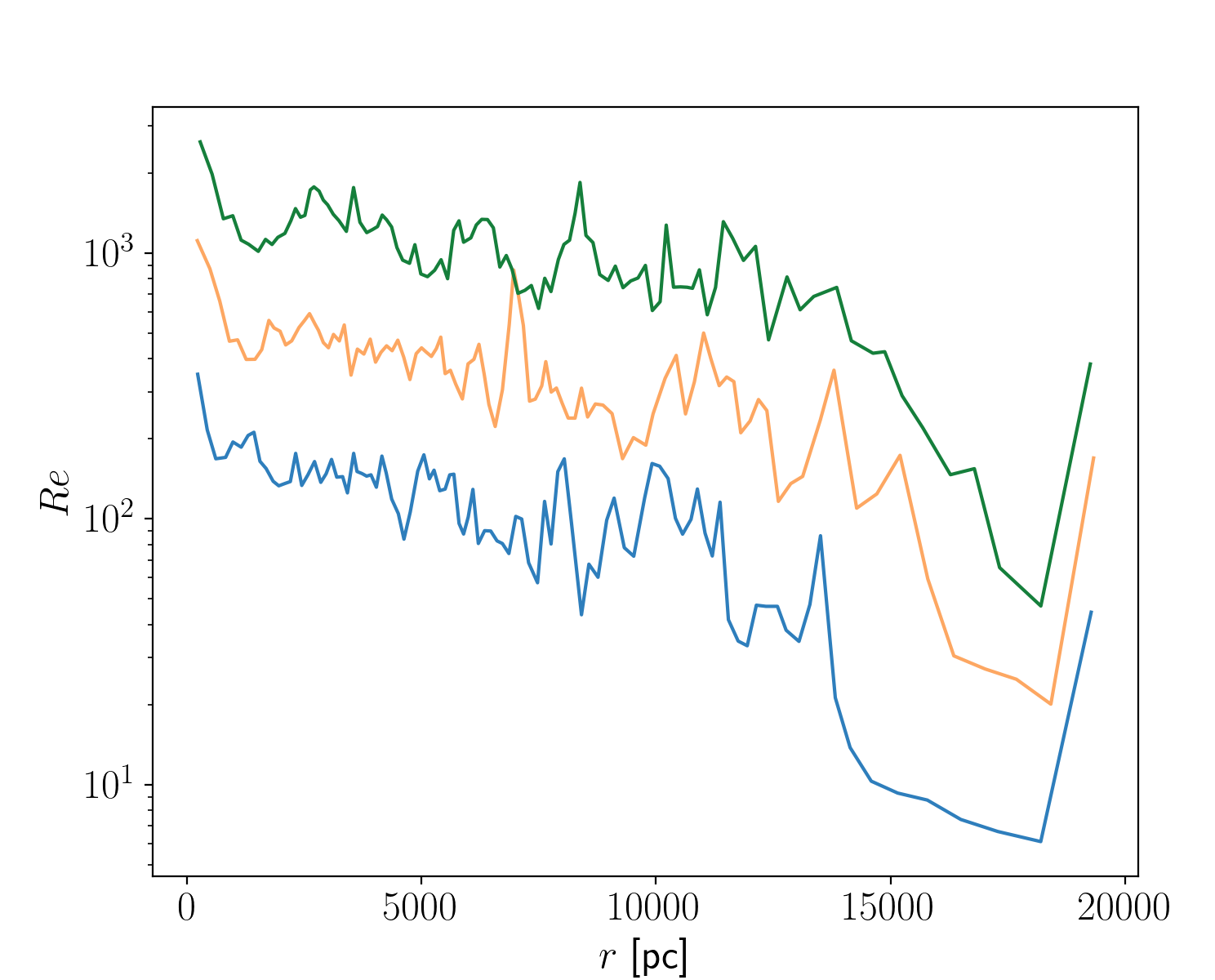}
    \includegraphics[width=0.9\hsize]{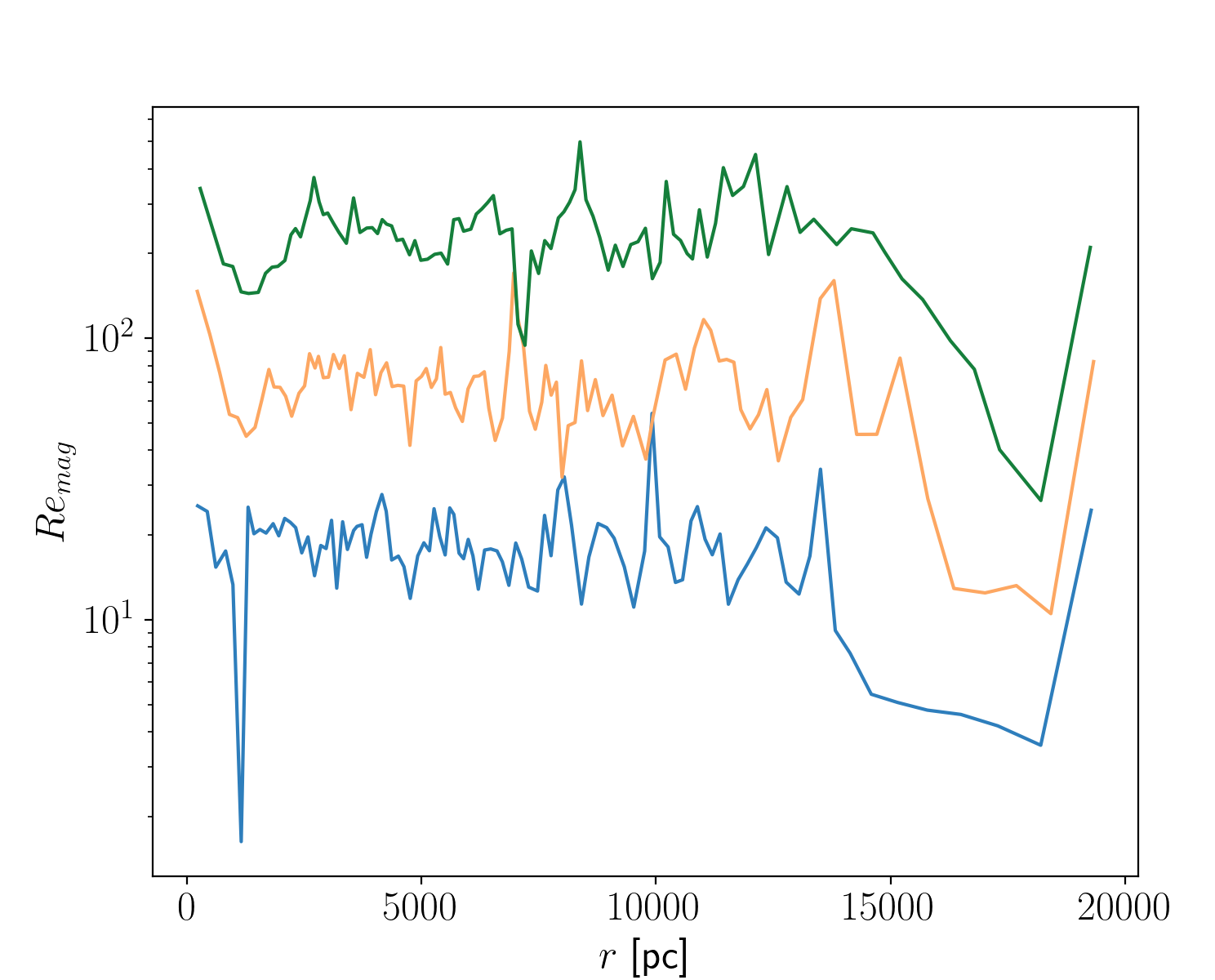}
    \caption{The density averaged radial profile of the fluid parameters for different resolutions at an early time ($t=250Myr$ $\alpha_B=0.5$) . From top to bottom we have: the Prandtl number, the magnetic Reynolds number and the regular Reynolds number. The Prandtl number seem to be largely independent of the resolution and mainly be effected by the conditions within the galaxy (supersonic + shear + stratified environment). Both the magnetic and regular Reynolds number can be seen to increase with a factor of $4$ as we increase the resolution, which is in agreement with the expected second order of the numerical diffusion}
    \label{fig:20}
\end{figure}
\\ \\
The time evolution of the central disk of all the 'Low' resolution runs can be seen in the left panel of \fig \ref{fig:15}. For the BW model, we can see that as we increase the feedback energy, we get less magnetic field amplification within the disk. For the SB model the amplification is independent of the feedback strength and the only case that see significant amplification is the simulation with $N_{FB}=200$. There is an interesting spike in the magnetic field strength for the $N_{FB}=200$ simulation early in its evolution. This is caused by the merger of two fragments within the central region of the disk. Significant amplification occurs within the shear layer between the two fragments during inspiral. The amplified magnetic field within the shear layer is subsequently diffused throughout the bulk and envelope of the two fragments. This is similar to what have been seen in high resolution simulations of binary neutron star mergers \citep{2021arXiv211208413P}. After the merger the produced magnetic fields are quickly diffused and advected away from the central region of the disk. This can be seen to occur in both the radial and the vertical directions, leading to a reduction in the mean magnetic field strength of the central disk.
\begin{figure*}[!ht]
    \centering
    \includegraphics[width=14cm]{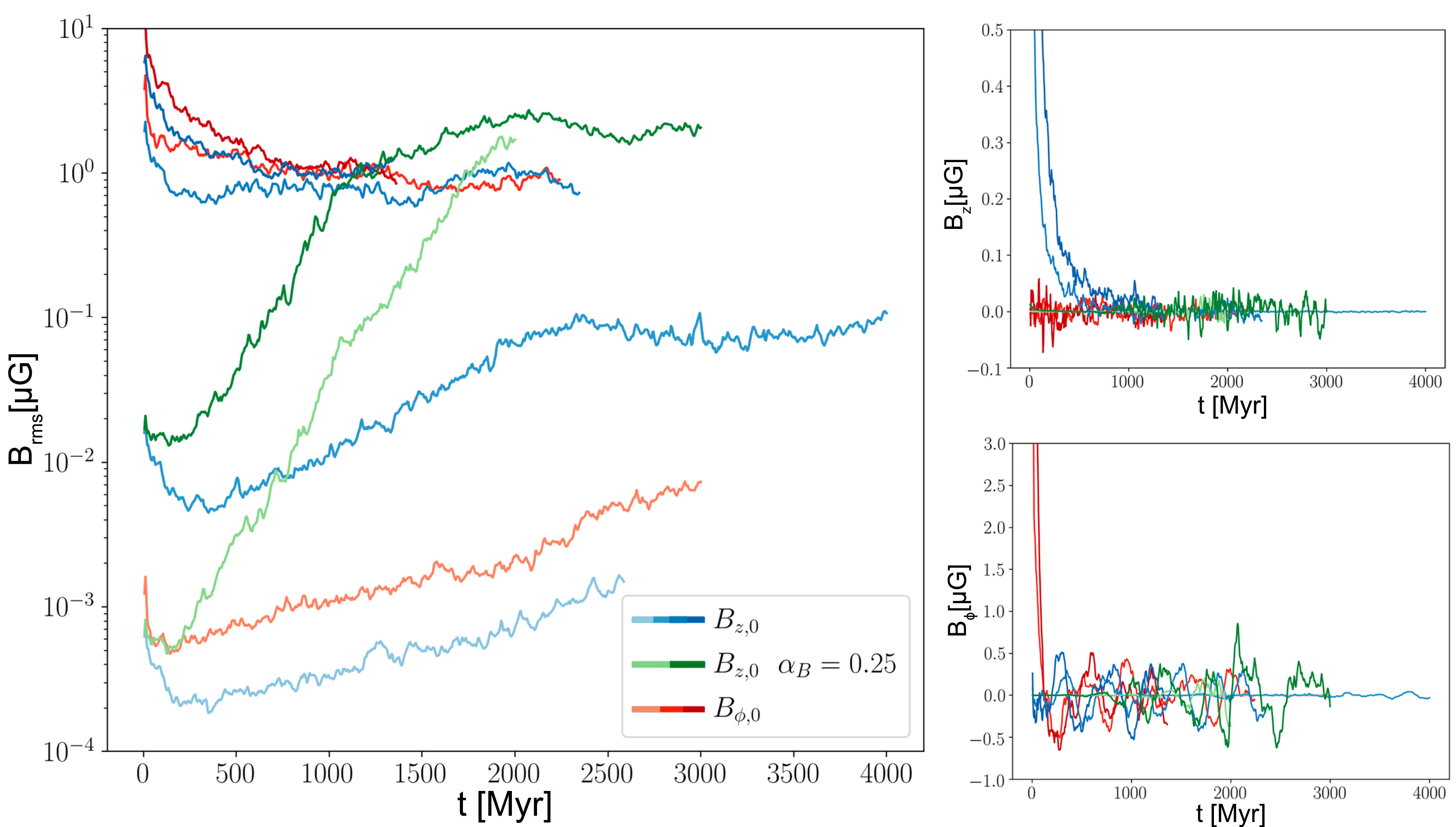}
    \caption{Time evolution of $B_{rms}$ (left panel) and the vertical and toroidal fluxes (right panel) within the central disk for different initial field strengths and geometry. Darker lines represent a lower initial thermal plasma beta $\beta_{th}$ within the centre of the disk. The distinct levels of initial central plasma beta are ($\beta_{0,centre}=(10^7,10^4,1,0.1)$). We can see that for the low plasma beta cases ($\beta_{0,centre}\leq 1$) the magnetic field strength is quickly reduced at the beginning of the simulation, but eventually saturates at a level around $1\mu G$. This is a similar saturation level as the $\alpha_B=0.25$ simulations that grow from much weaker initial fields at the same resolution. Due to this outflow of magnetic flux, we can see that all simulations reduce to a similar oscillating flux within the central disk for the vertical and toroidal fields.}
    \label{fig:21}
\end{figure*}
Looking at the right panel of \fig \ref{fig:15} we can see the time evolution of the 'Medium' resolution runs. Here we can see that all runs are amplified significantly and saturate on the order of $B=\propto 1\mu G$. Similar to the 'Low' resolution case, there is significant faster amplification in the case of the BW model, which reaches $B=\propto 1\mu G$ in about $500 $ Myr \footnote{Due to the average here being a simple mass average, the blastwave will be slightly biased due to having stronger magnetic field amplification in the inner regions.}. For the SB feedback we can see that $N_{FB}=64$ with $E_{SN}=0.5\epsilon_{SN}$ and $E_{SN}=1\epsilon_{SN}$ reaches saturation at about the same time. With higher feedback strengths we can see that it takes about $1500$ Myr to reach saturation within the central disk. With $N_{FB}=1$ saturation is reached after $1200$ Myr and the saturation level is lower and similar to that of the BW model.
\begin{figure*}[!ht]
    \centering
    \includegraphics[width=14cm]{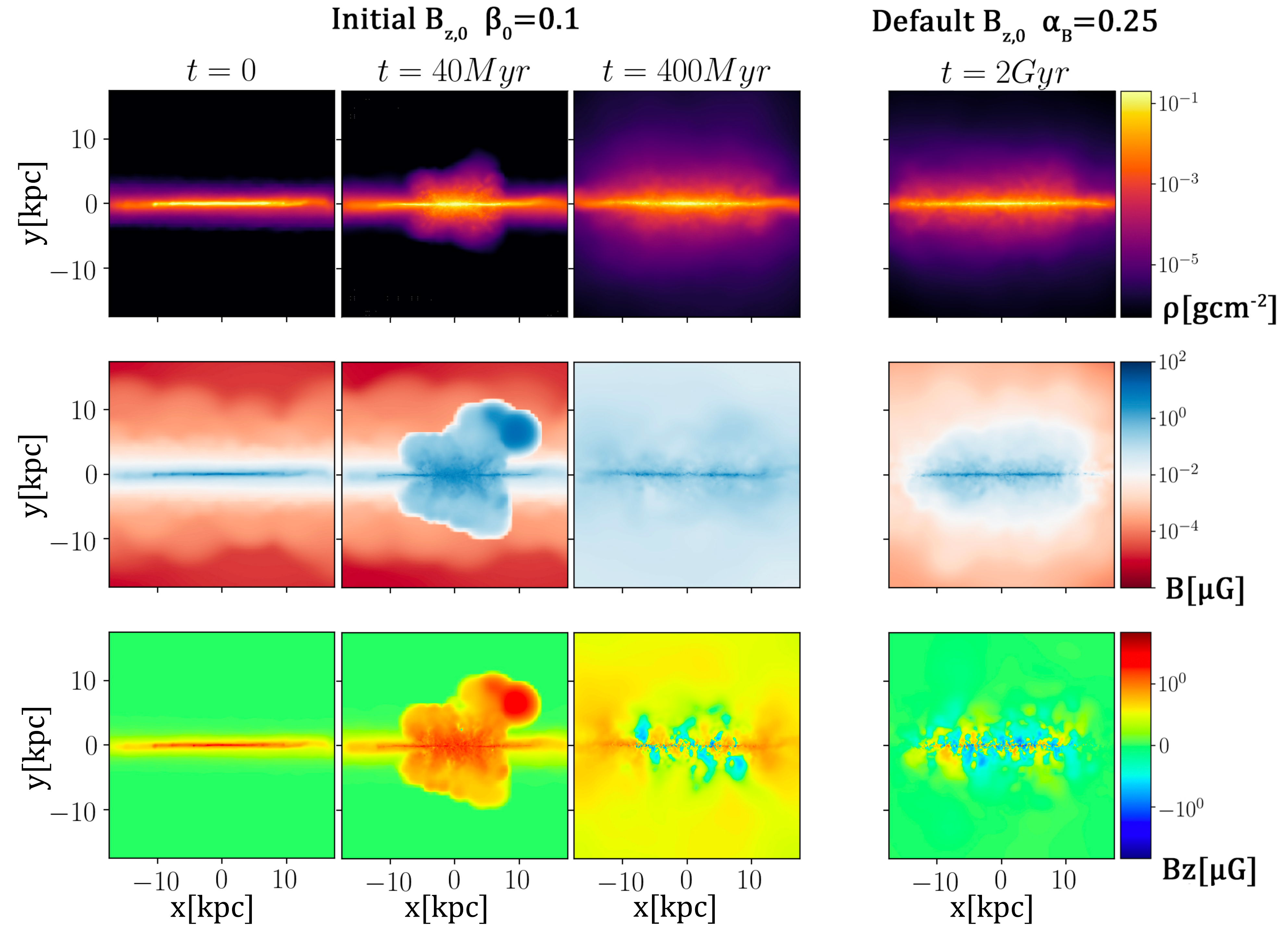}
    \caption{Rendering of density, magnetic field strength, and vertical magnetic field of the early starburst period for the initial low central plasma beta simulation ($\beta_{0,centre}=0.1$), which shows the ejection of magnetic flux during this time. This results in a more magnetized CGM than a simulation that amplifies from a much weaker initial magnetic field. However, both the cases can be seen to develop a similar vertical structure around the central disk }
    \label{fig:22}
\end{figure*}
\\ \\
The reasons for these differences are best illustrated by taking a look at the renders of the 'Medium' resolution runs in \fig \ref{fig:16}. Both the BW model and the SB model with $N_{FB}=1$ pushes material far away from the disk in the vertical direction, this allows for the advection of magnetic fields far from the disk. However, from the density rendering it appears that the SB model distributes gas closer to the disk than the BW model \footnote{Although the number of feedback particles for the BW model is in principle determined by the analytical calculation of the blastwave radius, in practise majority of feedback events has $N_{FB}=1$ for the resolutions considered in this work}. This is consistent with previous studies, which found that the SB model generally produce slower winds, but with higher mass loading factor \citep{2015MNRAS.453.3499K,2021A&A...655A..22M}. This implies that the turbulence generated within the ISM of the BW model is efficient in driving the dynamo, but remains not too disruptive to dampen the field. In comparison, the SB model with $N_{FB} =1$ disturbs the ISM more, leading to less amplification. The quick amplification in the BW model is likely further amplified by its higher star formation rate early on in the simulation, with a higher peak during the initial starburst and higher than all the SB models until around $350$ Myr. At the end of the simulation the star formation rate in the BW model is lower than the SB models. Increasing $N_{FB}$ from $1$ to $64$ essentially leads to more "gentle" galactic winds, with gas ejected being distributed closer to the disk. This increases the amount of resolved large-scale eddies in the vertical-direction, and thus we see a more effective dynamo.  Within the same model, increasing the feedback strength increases both the vertical and radial extent of the disk. In the bottom panel of \fig \ref{fig:16}, we can see the toroidal magnetic field in the disk. From this we can see that the scale of the toroidal mean-field increases with the feedback strength. How vertically concentrated the disk structure is seem to be a strong determinant of the amplification rate within the central disk seen in \fig \ref{fig:16}, even though the saturation level is higher in the SB models ($N_{FB}=64$). However, the lag in amplification of the higher feedback strength models may simply be due to these models having a larger volume, where more magnetic energy has to be produced in total before saturation can be reached.
\\ \\
Looking at the averaged vertical profile of the galaxy\footnote{Vertical profile from cylinder with a radius of $20$kpc around the centre of the galaxy.} in \fig \ref{fig:17}, it is clear that the energy density of the magnetic field is correlated to the turbulent energy density. The turbulent energy density of the BW model is significantly lower than the SB models close to the central disk. However, both profiles of the BW model and the $N_{FB}=1$ SB model remain more flat at large distances than the ones from simulations with $N_{FB}=64$. This is reflected in the magnetic energy density and is related to the advection of magnetic fields from the central plane. In the BW model we can see that the thermal energy density is significantly higher than the resulting turbulent energy close to the disk. This differs from the SB model, in which they are always closely linked. In general a lower turbulent velocity and lower mach number is seen in the BW model (see \fig \ref{fig:4}).
\\ \\ \\ \\
It is clear from these results that the 'Low' resolution simulations do not show the same converged behavior as the 'Medium' resolution. This points to a failure to resolve the relevant amplification processes in the 'Low' resolution simulations, either due to reduced dynamo efficiency or too much diffusion. In the next section we will have a closer look at this.
\subsubsection{Effect of Diffusion parameters}
\label{sec:diffsims}
Another important property in the amplification of the magnetic field comes from the diffusion parameters of the fluid: the Reynolds number, magnetic Reynolds number and the Prandtl number. In numerical simulations the two factors that determine these properties are the resolution and the numerical diffusion. The equations of the numerical estimated physical dissipation parameters are given in \sect \ref{sec:numdiff}. In these simulations we leave the numerical viscosity alone and only vary the numerical resistivity. This is done by changing the $\alpha_B$ parameter within the numerical scheme\footnote{Halving the $\alpha_B$ parameter leads to a halving of the numerical resistivity essentially.}. Care is however required, as a too low value can lead to excessive numerical noise/errors.
\\ \\
In \fig \ref{fig:18} and \ref{fig:20}, we compare four simulations, with different resolutions and different $\alpha_B$; two with $\alpha_B=0.25$ and two with the default value of $\alpha_B=0.5$. In \fig \ref{fig:18} we can see that lowering the numerical resistivity enables the "Low" resolution simulation to grow its field much faster, leading to saturation after about $1.9$ Gyr. Showing a similar resolved behaviour as the "Medium" resolution simulations (see also \fig \ref{fig:16}). Faster growth is also seen in the 'Medium' resolution ($\alpha_B=0.25$), where saturation is achieved after about $700$ Myr compared to $1100$ Myr in the default run. In addition, the field saturates at a higher level with $\alpha_B=0.25$. In \fig \ref{fig:19} (early time $t=0.25Gyr$) and \fig \ref{fig:20} (late time $t=2Gyr$) we can see the radial profile of the density averaged numerical dissipation parameters. Here, we have assumed that $L_{inj}=1kpc$ and $\sigma_{v}=v_{turb}$. The average Prandtl number can be seen to be below $1$ throughout the disk, with decreasing values towards the centre. Reducing the numerical resistivity can be seen to increase the Prandtl number and the magnetic Reynolds number as expected. At late times ($2Gyr$) we can additionally see that the Prandtl number increases with the resolution. This might be unexpected at first glance as both dissipation schemes are of second order ($\propto h^2$), that is to say that a doubling of resolution should result in 4 times lower numerical dissipation. This can be seen to be the case for the early time ($0.25Gyr$) simulations, where the Prandtl number stays roughly the same and the magnetic and regular Reynolds number increases equally in proportion to the resolution. This is because the density, velocity and magnetic field structure of the disk is more different between resolutions at later times than at early times. Another interesting behaviour of the Prandtl number is its decline towards the centre, as you would expect an equal or higher value in the central region. This is related to the curves of the magnetic and regular Reynolds number, where the magnetic Reynolds number has a more flat curve than the regular Reynolds number. We believe that this is related to the differences in artificial switches for the viscosity and resistivity. Artificial switches are based on local environment factors, to reduce the dissipation away from shocks. Meaning that the two schemes are not necessary correlated when the environment changes, which is what occurs when we move radially inward throughout the disk. The magnetic Reynolds number is shown here to be relatively low ($Re_{mag}=10-400$) compared to the levels potentially required for the small-scale dynamo ($Re_{mag}=30-2700$).
\\ \\
\subsubsection{Effect of initial magnetic field strength and geometry}
\label{sec:initmageffecsims}
For our "Low" resolutions simulations, we also explore what the effect of the initial magnetic field strength and geometry has on the evolution of the field. The initial magnetic field strength is varied, such that the central strength is equal to a set plasma beta value. Here we set the initial central plasma beta value to roughly $\beta_{0,centre}=(0.1,1,10^4,10^7)$, this represents a field strength of $B_{init}=(10,3,0.3,10^{-3}) \ \mu G$ at $\rho=\rho_0=6.77331\cdot 10^{-23} \ \rm g \ \rm cm^{-3}$. We simulate with both an initial vertical field and a toroidal field following \eq \ref{eq:vertfield} and \ref{eq:torfield}. For $\beta_{0,centre}=10^4,10^7$ we also vary the artificial resistivity parameter $\alpha_B=0.5$ and $\alpha_B=0.25$.
\\ \\
The time evolution of the magnetic field within the central disk for these simulations is plotted in \fig \ref{fig:21}. For the low plasma beta runs ($\beta_{0,centre}=0.1,1$), we can see that the magnetic field strength is quickly reduced at the beginning of the simulation but eventually saturates around $1\mu G$. This is just slightly lower than the saturation levels that are achieved by the lower resistivity runs ($\alpha_B=0.25$). For the high beta runs $\beta_{0,centre}=10^4,10^7$ we can see a similar behavior as the previous "Low" resolution simulations (\sect\ref{sec:resfb}-\ref{sec:fbsims}), where the effective amplification is highly damped and the "convergence" behaviour seen in the "Medium" and $\alpha_B=0.25$ "Low" resolutions is lost. There seems to be no real significant difference between starting with toroidal or vertical fields on the subsequent amplification. Looking at the right panel of \fig \ref{fig:21} we can see the evolution of the vertical and toroidal mean-fields. From this, we can see that the simulations with low plasma beta ($\beta_{0,centre}=0.1,1$) quickly lose their initial vertical/toroidal flux. In addition, we can see that the vertical flux simply reduces to noise around 0, while the toroidal flux can be seen to experience larger oscillations within the central disk.
\\ \\
The reason for the rapid flux removal of the low beta runs is shown in \fig \ref{fig:22}. Here we can see that the initial burst of feedback, blows the magnetic flux out from the central disk. Comparing this to the low resistivity run ($\alpha_B=0.25$) that amplifies from a low initial magnetic field strength, we can see that the biggest difference lies in the CGM. Where a strong vertical magnetic flux exists in the low beta simulation. The structure of the vertical magnetic field in the central disk, however, looks very similar between the two cases, confirming the evolution that we saw in \fig \ref{fig:21}. 
\section{Discussion}
\label{sec:discussion}
In this paper, we performed simulations of isolated Milky-Way type galaxies using SPMHD with a large range of different numerical parameters, such as supernova feedback, resolution, Jeans floor, diffusion parameters, and initial conditions. Looking at how each of these parameters affects the growth and saturation of the magnetic field.
\\ \\
Regions of the galaxy with an effective dynamo, saturate their magnetic fields with values ranging from $1$-$100\mu G$. The average saturation of the whole central disk lies around $2-5\mu G$, which is similar to the strength observed in the main body of the Milky Way disc and similar galaxies \citep{2009ApJ...702.1230T,2012ApJ...757...14J,2012ApJ...761L..11J,2016JCAP...05..056B}. This corresponds to an energy density at levels between $10$-$30\%$ of the thermal energy density. Increases in resolution and decreases in the numerical resistivity showed increased average saturation levels in the central disk, indicating non-convergence in the saturation level. Similar non-convergence was found for amplification rates, which continuously grew with increases in resolution. Simulations with feedback have saturation times in the ranges of $0.4Gyr$-$2Gyr$. For our simulations, we see no significant variation in the global star formation rates for galaxies with stronger magnetic fields. The saturation and growth rates are generally in agreement with past numerical simulations of MW-like isolated galaxies \citep{2013MNRAS.432..176P,2016MNRAS.457.1722R,2017ApJ...843..113B,2018MNRAS.473L.111S,2019MNRAS.483.1008S}. For our no-feedback runs with $N_J=8$, a 3-4 order of magnitude increase was seen for the magnetic field after around 500Myr, which is similar to the amplification seen by \cite{2009ApJ...696...96W} that did not include feedback in their runs. 
\\ \\
In most of our simulations, we see a decrease in magnetic field strength for the central region. This has been seen in previous galaxy simulations \citep{2009MNRAS.397..733K,2016MNRAS.461.4482D,2017MNRAS.471.2674R,2019MNRAS.485..117K}. There are, however, simulations that on the contrary produce very strong central magnetic fields \citep{2009MNRAS.397..733K,2013MNRAS.432..176P,2017ApJ...843..113B,2019MNRAS.483.1008S}.
In the case of \cite{2017ApJ...843..113B}, they include magnetic field injection during feedback events, which increases the magnetic field production within the central region of the galaxy. In addition, the galaxy that is simulated is relatively cold with very few outflows, making it more comparable to our no-feedback runs. For \cite{2013MNRAS.432..176P}, it was shown by \cite{2016MNRAS.463..477M} that this strong central amplification is removed when using the method of constrained transport to take care of the divergence error. The work of \cite{2009MNRAS.397..733K} and \cite{2019MNRAS.483.1008S} stem from the same SPH implementation of magnetic fields \citep{2009MNRAS.398.1678D}\footnote{While the magnetic field methodology is the same between the two works \citep{2009MNRAS.397..733K,2019MNRAS.483.1008S}, \cite{2019MNRAS.483.1008S} uses the updated GADGET3 SPH improvements of \cite{2016JCAP...05..056B} for their full MHD equations} and it is interesting to discuss the potential differences between our code and theirs. First of all, similar to \cite{2013MNRAS.432..176P}, the Powell method \citep{1999JCoPh.154..284P} is used to take care of the divergence \footnote{It is sometimes referred to as Powell cleaning, though there is no cleaning field introduced in this method. This is a zeroth-order cleaning as the method simply means removing the monopole currents from the induction equation, leading to the divergence errors to advect with the flow of the fluid (ensuring that the surface integral of the magnetic field is conserved \citep{2000JCoPh.160..649J,2001JCoPh.172..392D}.}. The magnetic diffusion within this code is done in two-part, first, a consistent magnetic diffusion is used, which is similar to ours, but with a different signal speed. The second diffusion operation done by the code is a kernel smoothing of the magnetic field every X timestep (X is a free parameter, but 15–20 in \cite{2009MNRAS.398.1678D}) to remove small-scale fluctuations, which is not a conservative operation. In \cite{2009MNRAS.397..733K} central disk amplification seems to be related to the divergence error, and we can reproduce this central amplification if we remove the divergence cleaning and reduce the artificial resistivity of our "Low" resolution runs. Though the field looks noisier than \cite{2009MNRAS.397..733K}, likely due to the neglect of the extra smoothing operation applied in these simulations. We find average divergence errors in the order of around $10^{-1}$, which are reduced with increased resolution. The increase in Jeans floor can also be seen to result in lower divergence errors, as it results in smoother structure and better-resolved gradients, that in turn causes less divergence errors to be produced. The divergence errors seen in our simulations are comparable or better than previous Lagrangian codes, which usually lie in the range $\epsilon_{div,B}=10^{-1}\leftrightarrow10^1$ \citep{2009MNRAS.397..733K,2013MNRAS.432..176P,2016MNRAS.461.4482D,2019MNRAS.483.1008S}. Another form for the divergence estimate was presented in \cite{2022ApJ...933..131S} ($\epsilon_{divB,i}=(\nabla \cdot B)_i \frac{B_i}{h_i} \sum_j \frac{h_i+h_j}{B_i+B_j}$), and the authors suggest that this form provides a more fair estimate of the error than the regular estimate of \eq \ref{eq:divBerr}. We disagree with this, as this added weighting does not make it dimensionless and biases the calculation giving a much lower divergence error than in actuality. It is also seen in some work that the mean of $\nabla \cdot B$ is used \citep{2013MNRAS.432..176P,2022ApJ...933..131S}, which is fairly redundant in Lagrangian codes that use the Powell method, as the total volume integral of $\nabla \cdot B$ across the simulation is constructed to be conserved in this method. In addition, the real measure of the $\nabla \cdot B$ error should be obtained by using the same gradient operator that is used in the magneto-hydrodynamic equations of the simulation. Using a higher order gradient estimator for this quantity than the one used in the induction equation of the simulation is likely to bias the analysis. 
\\ \\
It is clear that we have an active mean-field dynamo in many of our simulations. Within the gravitational unstable no-feedback runs this mean-field dynamo acts strongly in the filamentary structure between the generated fragments. The developed structure in the disk depends strongly on the cooling and the Jeans floor. For an effective amplification of the magnetic field, we find a "goldilocks zone" in the Jeans floor where the disk fragments but retains enough interconnectivity between the generated fragments to active the dynamo. This is similar to the recently described GI-dynamo process \citep{2019MNRAS.482.3989R}, where amplification occurs due to the vertical velocity rolls generated during spiral arm compression, which was found to be strongly dependent on the cooling. In \cite{2019MNRAS.482.3989R}, the authors find that amplification is hindered when the magnetic Reynolds number is increased above 100, due to the small-scale structure being generated within the spiral arms. In our case, the magnetic Reynolds number is generally below this for our no-feedback simulations. It would be interesting to run a no-feedback case at a much higher resolution to see if we also see this small-scale structure form, and if it can dampen the generated mean-field. 
\\ \\ 
The addition of feedback changes the amplification processes, the spiral arm structure is continuously disturbed by the feedback and there can be a significant transfer of magnetic flux outwards to the CGM. We found that, with sufficient resolution (high enough magnetic Reynolds number), the dynamo acts effectively in the disk given that the Jeans floor is small enough to not excessively suppress the collapse of gas into stars (removing the effect of feedback). For the magnetic field within the disk, the amplification is faster when the feedback is less disruptive to the ISM, and this is shown in the comparison between the blastwave and superbubble feedback models. The blastwave model generates hotter, faster winds that leaves the galaxy, and thus a relative thin, smoother disk, whereas the superubble model has a larger mass-loading and generate more ``fountain'' motions close to the galaxy, and thus increase the turbulence in the ISM and thicker disk. As a result, the blastwave model exhibit a faster amplification. However, the radial extent and maximum amplification level appear to be higher in the superbubble models in the converged runs (see Figure \ref{fig:16}), likely because it provides a more sustained injection of vertical motions, which enhances the dynamo. It is clear from Figure \ref{fig:17} that the magnetic energy density is correlated with the turbulent density. This is in agreement with the analytical behavior of the $\alpha\Omega$ dynamo \citep{1981SvA....25..553R,1988ASSL..133.....R,2005PhR...417....1B}, where the radial extent of the magnetic field increases with increasing disk scale-height. Nevertheless, we note that increased turbulence also enhance the diffusion, leading to more dissipation. In general, in a complex system like the galactic disk, the dependencies of dynamo processes on feedback can be highly non-linear. To further disentangle the dynamo processes active in these simulations would require a full mean-field analysis, which we leave for future work. 
\\ \\ 
As our simulations are isolated disks and does not have a realistic CGM component in the initial condition, the magnetization of the CGM is largely through the advection of magnetic field by galactic winds. Faster winds generally produce a larger magnetic field in the distant CGM (e.g. the blastwave model). However, the CGM closer to the galaxy sees higher magnetic field in all runs with slower winds. This can simply be because there are more gas closer to the galaxies in these models, but it is also very likely that local amplification occurs in the CGM, which in turn enhances dynamos within the disk plane, as indicated in our highest resolution runs (Figure \ref{fig:9}). In nature, the CGM in a cosmological environment also includes strong accretion flows which interact with galactic outflows, together with instability processes, where the magnetic field may play an important role as well \citep{2015MNRAS.449....2M}. In this regard, cosmological simulations are perhaps more appropriate to study the CGM. However, as the CGM are generally much less resolved than the disk, it is unclear whether dynamos in the CGM can occur in state-of-the-art cosmological runs (which have mass resolution close to our ``Medium'' resolution cases). We defer a more detailed study of the magnetic field in the CGM in future work.
\\ \\
From the simulations, we could see that the density averaged numerical Prandtl number is found to be below unity throughout the galaxy for all our simulations, with an increasing value with radius. It is clear that the Prandtl number is highly dependent on the underlying fluid environment. Early times ($0.25$ Gyr) in the galaxy evolution, the Prandtl number seems to be fairly independent of the resolution, while later times ($2$ Gyr), it shows a slight increase with resolution.  Previous studies done in shearing boxes with subsonic flow using the same code have shown a slight increase in the Prandtl number with resolution \citep{2021arXiv210501091W}. In \cite{2016MNRAS.461.1260T}, they find that the Prandtl number decrease with resolution in supersonic turbulent box simulations\footnote{Calculation of dissipation parameters are done in the continuum limit in this paper, which might effect the behavior of the Prandtl number}. An interesting follow-up work would be to further look into how the numerical Prandtl number changes as we go from subsonic to supersonic for different environments and to potentially modify the numerical scheme to produce a higher and more independent Prandtl number. This would be highly desirable as the Prandtl number can determine the growth and saturation level of many dynamo processes \citep{2004ApJ...612..276S,2014ApJ...797L..19F,2021arXiv210501091W}. The magnetic Reynolds number is within the range of ($Re_{mag}=10-200$) for all the simulations, which is comparatively low compared to the levels potentially required for the small-scale dynamo ($Re_{mag}=30-4000$). However, this depends on the assumptions of the turbulence driving length and the velocity dispersion at that scale, which we have taken to be $l_{inj}=1$ kpc and $\sigma_{v,inj}=v_{turb}$, respectively.
\\ \\
Fourier analysis was performed for a $10$ kpc cube region of the galaxy in order to investigate the relative solenoidal to compressive ratio for the velocity field (\eq \ref{eq:solratio} with $A=v$). We find that, assuming a turbulent injection scale of $1 $ kpc, we get a natural mixture value between $E_{kin,ratio}=0.5$-$0.65$ across our simulations, which is in accordance with the predicted natural mixture of 3D turbulence \citep{2004ARA&A..42..211E,2008ApJ...688L..79F}. While this indicates that we resolve the energy transfer between compressional and solenoidal modes in our simulations, it is too early to draw this conclusion due to the range of inertial scales and driving scales covered within the simulation. Local simulations looking at the turbulence driving length and mixture of solenoidal to compressive modes remain untested for the blastwave and the superbubble model. These would be important to get a better grasp of what the critical Reynolds number is for the small-scale dynamo of these models. Ideally, this would be tested using different boundary conditions as well (periodic, open, shear boundaries), as we know that flow conditions such as shear and vertical motions will affect both the active dynamo processes.
\\ \\
\section{Conclusions}
In conclusion, we find that  
\begin{itemize}
\item The results show a strong mean-field dynamo occurring in the spiral-arm region of the disk, related to an alpha-omega type dynamo, either by the classical alpha-omega effect or the recently described GI-dynamo.
\\ \\
\item Without star formation and feedback, the amplification is highly determined by the cooling and the smallest collapse-length set by the Jeans floor. As this determines the degree of fragmentation within the disk. The highest amplification can be seen in the case of $N_J=8$ for our 'Low' resolution runs. Amplification is driven by shear and vertical motions within the filamentary structure that forms around and between fragments ($\alpha\Omega$-effect).  Higher $N_J$ generally leads to less collapse and less amplification, and lower $N_J$ leads to too much fragmentation, decreasing the filamentary structure between fragments.
\\ \\
\item The inclusion of feedback is seen to work in both a destructive and positive fashion for the amplification process. Destructive interference for the amplification occurs due to the increase of turbulent diffusion within the disk and the ejection of magnetic flux from the central plane to the CGM. The positive effect of feedback is the increase in vertical motions and the turbulent fountain flows that develop. The effective amplification is highly dependent on the small-scale vertical structure and the numerical dissipation within the galaxy. Making the amplification highly dependent on resolution and the numerical dissipation parameters. Galaxies with an effective dynamo, saturate their magnetic energy density at levels between 10-30\% of the thermal energy density.
\\ \\
\item For the same feedback model and injection length, the amplification rate within the central disk reduces for stronger feedback (higher $\epsilon_{SN}$) runs, mainly due to an increase in scale height of the galaxy. This, however, leads to higher saturation of magnetic fields within the CGM, which can be shown to be directly correlated to the increase in turbulent energy density of the CGM. Given that the resolution is high enough, the saturation level within the central disk remains fairly independent of the feedback strength. 
\\ \\
\item It is clear from our results that the 'Low' resolution simulations do not show the same converged behavior as the 'Medium' resolution.  This points to a failure to resolve the relevant amplification processes in the 'Low' resolution simulations, either due to reduced dynamo efficiency or too much disruption and diffusion within the central disk.
\\ \\
\item Increasing the injection length of the superbubble feedback can be seen to have a positive effect on the magnetic field amplification. Potentially this is due to three effects, first, it can be seen to produce larger bubble regions that would indicate a larger turbulent injection length, which results in higher effective Reynolds numbers. Second, gradients become smoother and thereby more resolved. Third, the more "gentle" galactic winds produced at higher injection lengths distribute gas closer to the disk, leading to more large-scale eddies being resolved in the vertical direction. 
\\ \\
\item The blastwave scheme produces faster amplification than the superbubble scheme. The reason for this is harder to distinguish, however, we can see that blastwave generates a more compact and smooth inner structure where the majority of the amplification takes place. Thus, it can reduce turbulent diffusion and dissipation of ithe magnetic field. The blastwave model also produces hotter winds than the superbubble model, these winds will be fast but contain much less mass than the winds in the superbubble model. This leads to a reduced rate of advection of magnetic fields away from the central disk region, compared to the superbubble model. 
\\ \\
\item Due to the strong initial starburst within the galaxy the initial magnetic flux is ejected from the central disk to the CGM. This makes the subsequent evolution independent of the magnetic field geometry, where similar growth was seen for both an initial vertical field and an initial toroidal field. Stronger initial field strengths will disperse their initial flux from the central disk and saturate at a level comparable to galaxies that resolve the amplification process and amplify from a much weaker initial field strength.
\\ \\
\item The density averaged numerical Prandtl number is found to be below unity throughout the galaxy for all our simulations, with an increasing value with radius. During the early starburst period of the galaxy, the Prandtl number seems to be fairly independent of the resolution, while at later times ($2 Gyr$) we can see an increase in its value due to resolution. Previous studies have shown that in subsonic flow, the Prandtl number increases with the resolution for SPH \citep{2021arXiv210501091W}. Indicating a change in behavior for supersonic shearing flows. The magnetic Reynolds number is within the range of ($Re_{mag}=10-200$), which is comparatively low compared to the levels potentially required for the small-scale dynamo ($Re_{mag}=30-4000$). However, this depends on the assumptions of the turbulence injection length and the velocity dispersion at that scale, which we have taken to be 1kpc and $v_{turb}$ respectively.
\end{itemize}

\section*{Acknowledgements}
 We thank James Wadsley and Benjamin Keller for providing the isolated disk initial conditions and for the insightful discussions during the project. The simulations were performed using the resources from the National Infrastructure for High Performance Computing and Data Storage in Norway, UNINETT Sigma2, allocated to Project NN9477K. We also acknowledge the support from the Research Council of Norway through NFR Young Research Talents Grant 276043.  

\bibliographystyle{aa}
\bibliography{references}

\end{document}